\documentclass[11pt]{article}
\usepackage{amsmath,amsfonts,amsthm,mathabx,mathrsfs,graphicx,bm,enumerate,fancyhdr,mathtools, multicol, amssymb, caption, adjustbox, multirow, listings, float row, graphicx}
\usepackage{enumerate}
\usepackage{natbib}
\usepackage{url} % not crucial - just used below for the URL 
\usepackage{authblk}
\usepackage[margin=1in]{geometry}
\usepackage{algorithm}
\usepackage{algpseudocode}
\usepackage{booktabs,arydshln}
\usepackage{adjustbox}
\DeclareMathOperator*{\argmax}{arg\,max}

\newcommand{\E}{\mathbb{E}}

\newcommand{\indep}{\perp \!\!\! \perp}
\newcommand{\PM}{PM$_{2.5}$}

\makeatletter
\def\adl@drawiv#1#2#3{%
        \hskip.5\tabcolsep
        \xleaders#3{#2.5\@tempdimb #1{1}#2.5\@tempdimb}%
                #2\z@ plus1fil minus1fil\relax
        \hskip.5\tabcolsep}
\newcommand{\cdashlinelr}[1]{%
  \noalign{\vskip\aboverulesep
           \global\let\@dashdrawstore\adl@draw
           \global\let\adl@draw\adl@drawiv}
  \cdashline{#1}
  \noalign{\global\let\adl@draw\@dashdrawstore
           \vskip\belowrulesep}}
\makeatother

\newtheorem{assumption}{Assumption}

\newcommand{\blind}{0}

\usepackage{abstract}
\date{}

\begin{document}

\newcommand\blfootnote[1]{%
  \begingroup
  \renewcommand\thefootnote{}\footnote{#1}%
  \addtocounter{footnote}{-1}%
  \endgroup
}
\def\spacingset#1{\renewcommand{\baselinestretch}%
{#1}\small\normalsize} \spacingset{2}
\renewcommand\Authfont{\fontsize{12}{14.4}\selectfont} % author font
\renewcommand\Affilfont{\fontsize{11}{10.8}\itshape} % affil font

%% for comment bubbles...
%%%%%%%%%%%%%%%%%%%%%%%%%%%%%%%%%%%%%%%%%%%%%%%%%%%%%%%%%%%%%%%%%%%%%%%%%%%%%%
\spacingset{1}
\if0\blind
{
  \title{\vspace{-2em} \spacingset{1} \Large Environmental Justice Implications of Power Plant Emissions Control Policies: Heterogeneous Causal Effect Estimation under Bipartite Network Interference}
  
  \author[1]{Kevin L. Chen}
  \author[1]{Falco J. Bargagli-Stoffi}
  \author[1]{Raphael C. Kim}
  \author[1]{Rachel C. Nethery}
  \affil[1]{Department of Biostatistics, Harvard T.H. Chan School of Public Health (Boston, MA)}
  \maketitle
} \fi

\if1\blind
{
  \bigskip
  \bigskip
  \bigskip
  \begin{center}
  \vspace{-2em} \spacingset{1}
  \Large Environmental Justice Implications of Power Plant Emissions Control Policies: Heterogeneous Causal Effect Estimation under Bipartite Network Interference
  \end{center}
  \medskip
} \fi

\medskip
\begin{abstract}
\spacingset{1}

Emissions generators, such as coal-fired power plants, are key contributors to air pollution and thus environmental policies to reduce their emissions have been proposed. Furthermore, marginalized groups are exposed to disproportionately high levels of this pollution and have heightened susceptibility to its adverse health impacts. As a result, robust evaluations of the heterogeneous impacts of air pollution regulations are key to justifying and designing maximally protective interventions. However, such evaluations are complicated in that much of air pollution regulatory policy intervenes on large emissions generators while resulting impacts are measured in potentially distant populations. Such a scenario can be described as that of \textit{bipartite network interference} (BNI). To our knowledge, no literature to date has considered estimation of heterogeneous causal effects with BNI. In this paper, we contribute to the literature in a three-fold manner. First, we propose BNI-specific estimators for subgroup-specific causal effects and design an empirical Monte Carlo simulation approach for BNI to evaluate their performance. Second, we demonstrate how these estimators can be combined with subgroup discovery approaches to identify subgroups benefiting most from air pollution policies without a priori specification. Finally, we apply the proposed methods to estimate the effects of coal-fired power plant emissions control interventions on ischemic heart disease (IHD) among 27,312,190 US Medicare beneficiaries. Though we find no statistically significant effect of the interventions in the full population, we do find significant IHD hospitalization decreases in communities with high poverty and smoking rates.

\vspace{0.25in}


\end{abstract}

\vspace{2em}

\thispagestyle{empty}

% \noindent%
% {\bf Keywords:}  causal inference, heterogeneous treatment effects, bipartite network interference, spillover effects, inverse propensity weighting, air pollution, policy evaluation, environmental justice
% \vfill

\setcounter{page}{0}
\spacingset{2}
\newpage
\section{Introduction}
\label{sec:intro}

The risk of air pollution to human health has been well-documented over the past few decades. Previous studies have linked airborne pollutants, most notably fine particulate matter (\PM), to adverse human health effects, including cardiovascular and respiratory disease, stroke, and death \citep{Samet2000,Tsai2003,Koken2003,Dominici2006,Nethery2020,Wu2020,Henneman2023}. Furthermore, there has been a wealth of research demonstrating the disparate air pollution burden suffered by marginalized and minority groups. Several studies, for example, have noted disparate exposures and effects of air pollution for certain racialized groups and low-income communities \citep{Zeger2008,Hajat2015,Kioumourtzoglou2015,Di2017,jbaily2022air,Josey2023}. 
Recently, the U.S. Environmental Protection Agency (EPA) has placed particular focus on populations found to experience disproportionately increased risk from PM$_{2.5}$ exposure, citing the need for additional research on the disparities in environmental exposure and related health burdens among populations ``\textit{with environmental justice concerns, with a specific emphasis on minority populations and tribal communities}''
\citep{usepa2022a}. This is in line with their directive to achieve environmental justice, which the EPA defines as ``\textit{fair treatment and meaningful involvement of all people regardless of race} [...] \textit{or income with respect to the development, implementation, and enforcement of environmental laws, regulations, and policies}''\citep{usepa2022b}.

Title IV of the 1990 Clean Air Act Amendments aimed for a reduction of 10 million tons in annual sulfur dioxide (SO$_2$) emissions with respect to 1980 levels \citep{Chestnut1997}. To achieve these reductions, considerable focus has been placed on the installation of flue gas desulfurization (FGD) equipment (i.e., ``scrubbers'') on coal-fired power plants, which are the largest source of SO$_2$ emissions in the U.S. \citep{Massetti2017}. SO$_2$ emissions from these power plants are also a contributor to secondary PM$_{2.5}$ formation, which has been attributed to 460,000 deaths in the US Medicare population from 1999-2020 \citep{Henneman2023}. Many FGD technologies were designed for 90\% or greater SO$_2$ removal, with significant advances in both performance and cost-effectiveness since their introduction \citep{Srivastava2001}. As a result, these scrubbers play an important role in meeting the various targets set forth by the Clean Air Act to improve human health.

Due to the significant monetary expense and labor force associated with the installation and upkeep of interventions such as FGD scrubbers, it is necessary to consider the costs and benefits of retrofitting any given power plant. For example, consider a high-emitting power plant near a community with limited air conditioning access and therefore heavily exposed to outdoor pollutants via raised windows. A scrubber installed at this power plant may have a greater positive impact on human health than one installed at a scarcely-used power plant near a neighborhood with high air conditioning access. Thus, for those designing air pollution interventions, it is important to understand how the characteristics of both the power plant and the potentially impacted population may influence the magnitude of the intervention's health impacts. 

The nature of air pollution and the complex mechanisms through which it reaches individuals present numerous methodological challenges to quantifying the health impacts of scrubber installation and similar air pollution interventions. Such challenges are exacerbated when accounting for the potentially heterogeneous impacts of the same intervention across population subgroups. Pollutants are emitted from, and any scrubbers are installed on, coal-fired power plants, while health effects are experienced by individuals and communities. This results in what is known as a bipartite structure in the data, where units on which the interventions are applied differ from the units on which outcomes of interest are observed. Moreover, air pollution transport--the complex process in which pollutants react in the atmosphere and are transported (e.g., by the wind)--means that intervening at a single power plant can potentially affect health in many distant communities, creating a dense network of connections between power plants and communities \citep{Zigler2020}. This is a variant of the phenomenon known as ``network interference'' in the causal inference literature (described below). The study of causal effects in this scenario has thus been described as \textit{bipartite causal inference with interference} \citep{Zigler2021}, and henceforth referred to in this paper as causal inference with \textit{bipartite network interference} (BNI).

The network structure and presence of interference violate the Stable Unit Treatment Value Assumption (SUTVA) commonly held in classical causal inference literature and studies of heterogeneous treatment effects (HTEs) \citep{rubin1986comment}. SUTVA requires that the treatment status of one unit does not impact the outcomes of other units. However, there are many real-world instances where the treatment assigned to one unit can spill to other units. In the presence of interference, the effect of the treatment status of other units on one's outcome is usually referred to as a \textit{spillover effect}. Ignoring or oversimplifying the interference structure in analyses could lead to biased treatment effect estimation\citep{Forastiere2020,Sobel2006}. Furthermore, detecting or estimating spillover effects is key to provide a complete understanding of the causal effects of a certain treatment \citep{bargagliclusterednetwork}. In our motivating problem, spillover effects are crucial to evaluating the efficacy of scrubbers due to air pollution transport. For instance, treating a power plant may have effects both on its immediate surroundings (i.e., direct effect) and on distant communities affected by its emissions (i.e., spillover effect).

In the non-bipartite setting, \cite{Hudgens2008} as well as \cite{Tchetgen2010} approach the interference problem using cluster and group-randomized designs, while \cite{Forastiere2020} propose an approach based on a modified neighborhood SUTVA and extended unconfoundedness assumption. In the bipartite setting, \cite{Zigler2021} formulate bipartite estimators in a partial interference setting, while \cite{Zigler2020} extend this work in a more general bipartite setting, providing estimators of population average treatment effects. To date, however, all previous work in the BNI setting has been focused on the estimation of sample/population average effects, and extensions of this existing methodology to estimate heterogeneous treatment effects and discover subgroups whose effects differ from the population average effects have not yet been explored. In the context of power plant emissions, this is a crucial gap in the literature, as highlighted by the EPA in its push to achieve environmental justice objectives \citep{usepa2022a}. This motivates the need for the development of statistical methods to study heterogeneous causal effects under the BNI setting.

In this paper, we innovate the literature in various ways. First, we propose causal estimators for subgroup-specific treatment effects (i.e., conditional average treatment effects) under BNI, relaxing assumptions of partial or no interference, via augmented inverse propensity weighting (AIPW), stabilized IPW, and G-computation methods. Second, we design and implement a novel empirical Monte Carlo simulation approach using a combination of real-world and simulated data to evaluate HTE estimators under BNI. Third, we demonstrate how these estimators can be combined with approaches for subgroup discovery to identify subgroups that benefit most from scrubber installations without a priori specification. Finally, starting from our motivating application, we apply the proposed estimators to estimate the overall and subgroup-specific effects of scrubber installation at coal-fired power plants on ischemic heart disease (IHD) hospitalizations among Medicare beneficiaries. These contributions are of particular importance in this context, as existing methodologies have not considered heterogeneous treatment effect estimation or data-driven subgroup discovery in the BNI setting. This novel methodology expands upon previous literature through careful consideration of the structures by which air pollution affects human health, in order to accurately inform policy that promotes equitable improvement in air quality and overall environmental health.

The reminder of this paper is organized as follows. In Section~\ref{sec:setup} we introduce the problem of bipartite network interference in causal inference, the identifying assumptions, and the causal estimands of interest, and then we detail our proposed methodology in Section~\ref{sec:methods}. Section~\ref{sec:simulations} details an extensive Monte Carlo simulation study for BNI through which we evaluate the performance of the proposed estimators. The results of our motivating application are presented in Section~\ref{sec:application}. Section~\ref{sec:discussion} concludes the paper.

%\section{Bipartite Network Interference and Causal Effects}
\section{Set Up}\label{sec:setup}

\subsection{Motivating Data and BNI Concepts and Notation}\label{sec:bipartite}

Our motivating application mirrors that of \cite{Zigler2020}, in which the aim is to estimate the impact of power plant scrubber installations on human health in a context where the complex interference structure between power plants and populations is driven by air pollution transport. However, no literature to date has looked at heterogeneity in the impacts of such policies. To characterize the interference structure, we utilize the HYSPLIT Average Dispersion (HyADS) model, a reduced-complexity pollution transport model that estimates ZIP code level emissions exposures from individual coal–fired power plants in the U.S. \citep{Henneman2019}. From HyADS, we obtain a source-receptor matrix, where each matrix element represents a unit-less metric of the impact that emissions from an individual power plant have on a particular ZIP code, hereafter referred to as ``HyADS scores''. Data on scrubber installations in the year 2005 and plant characteristics for 314 coal–fired power plants concentrated in the Eastern US, which serve as our intervention units of interest, were obtained from the U.S. Environmental Protection Agency Air Markets Program Database. The outcome units for our analysis are ZIP codes. Our health outcome of interest is the ZIP code-level ischemic heart disease (IHD) hospitalization rate in 2005 among Medicare beneficiaries. For each ZIP code, U.S. Census socioeconomic and demographic features (for the year 2000), meteorological \citep{Kalnay1996}, and smoking rate \citep{DwyerLindgren2014} covariate data were obtained. We wish to estimate the causal effects of scrubber installation on IHD hospitalization rates and study how these effects vary for different population subgroups.

BNI arises when interventions are implemented on one type of unit (e.g., power plants), while outcomes are measured on another, distinct, type of unit (e.g., people/communities), and the mechanisms behind how interventions affect outcomes lack a direct one-to-one mapping. We introduce the concepts critical to formalize the BNI setup here. Consider a set of intervention units indexed by $j=1,2,\dots,J$ on which treatments are potentially imposed. Then, let $i=1,2,\dots,n$ index another set of units on which the outcome is measured (outcome units), which is disjoint from the set of intervention units. We will consider a bipartite network structure connecting the two; more specifically, a weighted directed graph allowing for connections between the sets of intervention and outcome units, but not within either set. The weights of the connections between units are defined in a $J \times n$ matrix, $\mathbf{H}$, where the entry $h_{ji}$ is the strength of the effect of intervention unit $j$ on outcome unit $i$. Interference in this scenario, therefore, arises without a one-to-one mapping of connections in this network. In our motivating data, $\mathbf{H}$ is the HyADS source-receptor matrix with entries representing the degree to which emissions from each power plant affect each ZIP code.

In non-bipartite settings, when interference is present between units, units' treatments are often expressed as treatment vectors composed of the unit's own treatment along with the treatment statuses of other units affecting them. In the most general bipartite scenario, an outcome unit's treatment can be defined as the vector of treatments at all intervention units. Under the potential outcomes framework \citep{Rubin1974}, this setup presents an issue when considering causal effects, since the number of potential outcomes grows exponentially with an increasing number of intervention units (i.e., $2^J$ potential outcomes, where $J$ is the number of intervention units). This motivates the need for a reduction in the dimension of the treatment vector. In the bipartite setting, as well as the more general setting of network interference, studies have proposed a characterization of treatment effects as direct and indirect effects (aka spillover) through a simplification of treatment statuses \citep{Forastiere2020}.

\begin{figure}[ht!]
    \centering
    \includegraphics[width=0.8\textwidth]{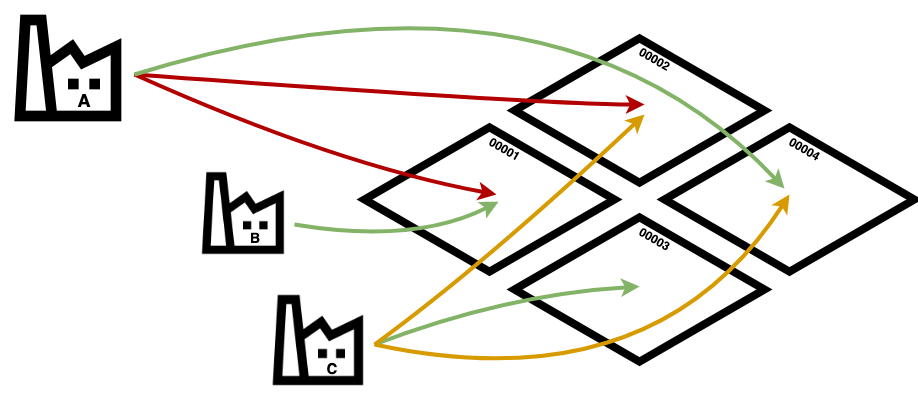}
    \caption{A simplified example of a bipartite network. In this example, power plants A, B, and C are intervention units, and ZIP codes 00001, 00002, 00003, and 00004 (depicted by diamonds) are outcome units. Arrow colors represent the strength of the influence that the power plant has on the ZIP code.}
    \label{fig:bipartiteexample}
\end{figure}

In the air pollution transport scenario, previous literature has condensed the outcome units' treatment information into two dimensions as follows: (1) a ``key associated'' treatment, which is based on the treatment of a single intervention unit, and (2) an ``upwind'' treatment, which is a function of some/all other intervention units' treatments arising from interference \citep{Zigler2021}. In our motivating application, the units of intervention are power plants and the outcome units are ZIP codes. The key-associated treatment variable, denoted $Z_i \in \{0,1\}$, typically reflects the treatment status of the most influential power plant on a ZIP code (i.e., the treatment of power plant $j$ where $j = \argmax_j h_{\bm{\cdot} i}$ for ZIP code $i$). Thus, the power plant that has the most influence on a given ZIP code is referred to as the key-associated power plant for that ZIP code. The upwind treatment can be defined in any number of different ways--e.g., most simply as the treatment status of the second-most influential power plant on a ZIP code. The upwind treatment idea relates to that of an \textit{exposure mapping}, which maps a treatment vector to some condensed exposure value \citep{Aronow2017}. To give a more general formalization of the upwind treatment status, let $\bm{T}_{\bm{\underline j}_{(i)}^\dagger}$ be a vector of the treatment status for all intervention units, except for the key-associated unit $j_{(i)}^*$. Then, $\gamma:\bm{T}_{\bm{\underline j}_{(i)}^\dagger} \to G_i\in\{0,1\}$ is some binary mapping of the treatment status of intervention units that are not key-associated with outcome $i$, and the resulting $G_i$ is the upwind treatment status. A simplified example diagram of the BNI setting in this context is shown in Figure \ref{fig:bipartiteexample}. 

\spacingset{1.5}

\spacingset{2}
\subsection{Causal Estimands and Identifying Assumptions} \label{sec:effects}

In the remainder of this section, we refer to units in the general framing of ``intervention'' and ``outcome'' units.

\subsubsection{Causal Estimands}

Let $Y_i$ be the observed outcome for outcome unit $i$, and let $\bm{X}_j^{int}$ and $\bm{X}_i^{out}$ denote covariates measured at the intervention unit and outcome unit levels, respectively. The preclusion of SUTVA in the BNI setting leads us to seek different assumptions to define the potential outcomes. We adapt the standard causal assumptions of overlap, consistency, and ignorability to the BNI setting (detailed in Appendix~\ref{sec:assumptions}), in addition to the following assumption, as previously defined in \cite{Zigler2020}:

\begin{assumption}[Upwind Interference]
Given $Z_i$ and $G_i$, if $Z_i=Z_i'$ and $G_i=G_i'$, then \[Y_i(Z_i,G_i) = Y_i(Z_i',G_i').\]
\end{assumption}
\noindent where $Y_i(z,g)$ is outcome unit $i$'s potential outcome under key-associated treatment $z$ and upwind treatment $g$. This assumption implies that the potential outcomes for a given outcome unit $i$ depend only on the full treatment vector $\bm{T}$ through the key-associated treatment $Z_i$ and upwind treatment $G_i$. 

Following \cite{Forastiere2020}, we can define average causal estimands in network interference settings as differences between mean potential outcomes. Similarly, we define expected potential outcomes as:
\[\mu(z,g) = \E\left[Y_i(Z_i=z, G_i=g)\right],\]
where $\mu(z,g)$ is the marginal mean of the potential outcome for an outcome unit $i$ under key-associated treatment $z$ and upwind treatment $g$. To extend this to the study of subgroup-specific treatment effects, we also define expected potential outcomes conditional on some outcome unit covariates, $x$, as
\[\mu(z,g; \; x) = \E\left[Y_i(Z_i=z, G_i=g) \mid \bm{X}_i = x\right].\]

We detail two estimands of interest using these average potential outcomes. The first illustrates the effect of changing the key-associated treatment status while holding the upwind treatment status constant, or the ``average direct effect'':
\[\tau(g) = \mu(1,g) - \mu(0,g).\]

The second estimand results from changing the upwind treatment status while keeping the key-associated treatment constant, or the ``average spillover effect'':
\[\delta(z) = \mu(z,1) - \mu(z,0).\]

In our air pollution application, the average direct effect can be viewed as the average effect of installing a scrubber on a given power plant, conditional on the installation status of all other power plants in the interference network. The average spillover effect can be analogously viewed as the effect of spillover (e.g., from pollution dispersion) from upwind power plants, conditioning on the scrubber installation status of the key-associated plant.

Similarly, the conditional average direct effect and conditional average spillover effect are respectively defined as
\[\tau(g; \; x) = \mu(1,g; \; x) - \mu(0,g; \; x),\]
\[\delta(z; \; x) = \mu(z,1; \; x) - \mu(z,0; \; x),\]
where the interpretation of these conditional mean effects is similar to their population average counterparts, except that they are conditional on covariates $x$.

\section{Novel Estimators under BNI}\label{sec:methods}
\subsection{G-computation}
\label{sec:estimators}

We start by defining a G-computation estimation method which will serve as a point of comparison. Let $\hat{\mu}_i(z,g;x^{int},x^{out})$ be unit $i$'s predicted outcome from some conditional outcome model. Then, the G-computation estimator for the average direct and spillover effect is defined as
\begin{equation}
    \hat{\tau}^G(g) = \hat{\mu}^G(1,g)-\hat{\mu}^G(0,g) \quad \text{and} \quad \hat{\delta}^G(z) = \hat{\mu}^G(z,1)-\hat{\mu}^G(z,0),
\end{equation}
where
\begin{equation}
    \hat{\mu}^G(z,g) = \frac{1}{n}\sum_{i=1}^n \hat{\mu}_i(z,g;x^{int},x^{out}).
\end{equation}

We now define a propensity score to predicate the development of propensity score-based estimators. Denoting $T_j$ to be the treatment corresponding to the intervention unit $j$, we define an intervention-level propensity score as
\begin{equation}
    \phi_j(t; \; \bm{x}^{int}, \; \bm{x}^{out}) \ = \ P\left(T_j=t \mid \bm{X}_j^{int} = \bm{x}^{int}, \; \eta_j(\bm{X}_{1:n}^{out}) = \eta_j(\bm{x}_{1:n}^{out})\right).
\end{equation}
% \rck{\begin{align*}
%     \phi_{[J],h,\eta}(t; \; \bm{x}^{int}, \; \bm{x}^{out}) \ &= \ P\left({ h_{tx}(T_{[J]}) }=t \mid h_{int}(\bm{X}^{int}) = h_{int}(\bm{x}^{int}), \; \eta(\bm{X}_{1:n}^{out}) = \eta(\bm{x}_{1:n}^{out})\right).
% \end{align*} 

% where $T_{[J]}$, is the treatment random variable, $\eta$ is some mapping of all outcome units, and $h=\{ h_{tx}, h_{int} \}$ is some transformation applied to the vector of treatment units $T_{[J]}$ and intervention units respectively. For example, $h_{tx},h_{int}$ can take the key associated treatment unit with some $i$ and key associated power plant as we'll see below. Other functions can specify how to transform treatment and intervention units such as the mean.

The probability $\phi_j(t; \; \bm{x}^{int}, \; \bm{x}^{out})$ is conditional on the intervention-level covariates of treatment unit $j$ as well as a function $\eta(\cdot)$ of the covariates for \textit{all} outcome units. We include outcome unit covariates in our intervention-level propensity score due to the potential for these covariates to affect treatment assignment. In our motivating problem, for example, we might expect factors such as population size or other demographic variables to play a role in a policymaker's decision whether or not to treat a power plant.
To conduct analyses, we map these propensity scores to the outcome unit level, based on each outcome unit's key-associated and upwind intervention units. This requires an additional assumption, outlined as follows:

\begin{assumption}[Treatment assignments are independent.]
For all $j,j' \in \{1,\dots,J\}$,
\[T_j \indep T_{j'} \mid \bm{X}_j^{int}, \; \bm{X}_{j'}^{int}, \; \eta_j(\bm{X}_{1:n}^{out}), \; \eta_{j'}(\bm{X}_{1:n}^{out})\]
\end{assumption} 
This assumption states that treatment assignments are independent, conditional on power plant covariates and respective functions of ZIP-level covariates. 

We adopt a joint propensity score approach similar to that of \cite{Forastiere2020}, resulting in a joint propensity score for each outcome unit defined as follows:
\begin{equation}
    \psi_i(z,\; g\; ; \; x^{int}, \; x^{out}) \ = \ \phi_{j_{(i)}^*}(z; \; x_{j_{(i)}^*}^{int}, \; x^{out}) \ \times \ P\left(\gamma \left(\bm{T}_{\bm{\underline j}_{(i)}^\dagger}\right)=g \ \mid \ \mathbf{X}^{int}_j, \ \eta_j(\mathbf{X}^{out}_{1:n}); \ j \in \bm{\underline j}_{(i)}^\dagger\right),
\end{equation}

where $\bm{\underline j}_{(i)}^\dagger$ denotes the set of intervention units included in the outcome unit $i$'s upwind treatment set and $\gamma$ is a mapping of the treatment values of units in $\bm{\underline j}_{(i)}^\dagger$. For example, as in Section~\ref{sec:bipartite}, $P\left(\gamma \left(\bm{T}_{\bm{\underline j}_{(i)}^\dagger}\right)=g \ \mid \ \mathbf{X}^{int}_j, \ \eta_j(\mathbf{X}^{out}_{1:n}); \ j \in \bm{\underline j}_{(i)}^\dagger\right)$ could be defined simply as the propensity score for the second-most influential power plant on outcome unit $i$.

% \rck{\[\psi_{i, \mathcal{Z}_i, \mathcal{G}_i, \eta}(z,\; g\; ; \; x^{int}, \; x^{out}) \ = \ \phi_{[J], \mathcal{Z}_i, \eta}(z; \; x_{}^{int}, \; x^{out}) \ \times \ \phi_{[J], \mathcal{G}_i, \eta}(g; \; x_{}^{int}, \; x^{out}) \]
% }

% \rck{where $ \mathcal{Z}_i=\{  \mathcal{Z}_{i,tx}, \mathcal{Z}_{i,int} \}$ takes the key associated mapping of $T_{[J]}$ for unit $i$ and key associated int unit for unit $i$, and $\mathcal{G}_i$ performs the analogous operation on the second key associated plant of unit $i$.}

\subsection{Population Average Effect Estimators via Augmented Propensity Score Weighting}
\label{sec:pate}
% Estimation
In the remainder of this section, we propose two propensity score-based estimators for the estimation of direct and spillover treatment effects in the BNI context. \cite{Qu2022} define similar doubly-robust estimators in a conditional exchangeability framework, where exchangeability is assumed to hold among subsets of units. However, to our knowledge, there are currently no comparable causal estimators in this setting that do not impose a clustering assumption on outcome units. First, we present BNI-specific AIPW estimators for the direct and spillover effects:
\begin{equation}\hat{\tau}^{AIPW}(g) = \hat{\mu}^{AIPW}(1,g)-\hat{\mu}^{AIPW}(0,g) \quad \text{and} \quad \hat{\delta}^{AIPW}(z) = \hat{\mu}^{AIPW}(z,1)-\hat{\mu}^{AIPW}(z,0),\end{equation}
where
\begin{equation}
    \hat{\mu}^{AIPW}(z,g) = \frac{1}{n} \sum_{i=1}^n \Bigg\{\frac{I\{Z_i=z, G_i=g\}}{\hat{\psi}_i(z,g;x^{int},x^{out})}\cdot Y_i + \bigg(1-\frac{I\{Z_i=z,G_i=g\}}{\hat{\psi}_i(z,g;x^{int},x^{out})}\bigg)\cdot \hat{\mu}_i(z,g;x^{out})\Bigg\}.
\end{equation}
$\hat{\psi}_i(z,g;x^{int},x^{out})$ is an estimate of unit $i$'s joint propensity score and $\hat{\mu}_i(z,g;x^{int},x^{out})$ is a prediction of unit $i$'s outcome from any conditional mean outcome model.

The multiplicative nature of the joint propensity score can result in extreme values of treatment propensities, particularly resulting in propensity scores near zero. This results in extremely large weights using IPW methods, in some cases leading to biased estimation and extreme variability \citep{Cheng2022}. Previous studies have shown that the use of stabilized weights improves on unstabilized IPW approaches by moderating these extreme propensity weights \citep{Robins2000,Chesnaye2021}. Thus, we also propose BNI-specific modified estimators with stabilized weights, which we call stabilized augmented inverse propensity weighting (SAIPW) estimators:
\begin{equation}\hat{\tau}^{SAIPW}(g) = \hat{\mu}^{SAIPW}(1,g)-\hat{\mu}^{SAIPW}(0,g) \quad \text{and} \quad \hat{\delta}^{SAIPW}(z) = \hat{\mu}^{SAIPW}(z,1)-\hat{\mu}^{SAIPW}(z,0),\end{equation}
where
\begin{equation}
\begin{split}
\hat{\mu}^{SAIPW}(z,g) &= \frac{1}{n} \sum_{i=1}^n \Bigg\{\bigg(\frac{1}{n}\sum_{k=1}^n \frac{I\{Z_k=z, G_k=g\}}{\hat{\psi}_k(z,g;x^{int},x^{out})}\bigg)^{-1}\frac{I\{Z_i=z, G_i=g\}}{\hat{\psi}_i(z,g;x^{int},x^{out})}\cdot Y_i \\
    &+ \bigg[1-\bigg(\frac{1}{n}\sum_{k=1}^n \frac{I\{Z_k=z, G_k=g\}}{\hat{\psi}_k(z,g;x^{int},x^{out})}\bigg)^{-1}\frac{I\{Z_i=z,G_i=g\}}{\hat{\psi}_i(z,g;x^{int},x^{out})}\bigg]\cdot \hat{\mu}_i(z,g;x^{int},x^{out})\Bigg\}.
\end{split}
\end{equation}
Truncation of propensity scores, which can be used either independently or in conjunction with stabilization, is recommended in general \citep{cole2008constructing} and may be especially useful here given the tendency of the joint propensity score to yield extreme weights.

\subsection{Heterogeneous Effect Estimators under BNI}
\label{sec:cate}

As we are also interested in the estimation of treatment effects among subgroups of the population defined by covariate values, we consider the use of G-computation and AIPW estimation to estimate HTEs \citep{Lee2017, Li2021}. In the remainder of the paper, we refer to conditional treatment effects as HTEs. We thus introduce estimators analogues of conditional average direct and spillover effects for some set of observed covariate values $x$: 

\begin{equation}\begin{gathered}\hat{\tau}^m(g;\;x) = \hat{\mu}^m(1,g;\;x)-\hat{\mu}^m(0,g;\;x), \\ \hat{\delta}^m(z;\;x) = \hat{\mu}^m(z,1;\;x)-\hat{\mu}^m(z,0;\;x),\end{gathered}\end{equation}

\normalsize

\vspace{1em}
\noindent where $m$ defines the particular method used (i.e., G-computation, AIPW, or SAIPW) and we plug in $\hat{\mu}^m(z,g;\;x)$ from the respective method, which are defined as follows:

\vspace{-1em}
\begin{equation}
    \hat{\mu}^G(z,g;\;x) = \frac{1}{n_{x}}\sum_{i:x_i=x} \hat{\mu}_i(z,g;x^{int},x^{out}),
\end{equation}

\vspace{-1em}
\begin{equation}
    \hat{\mu}^{AIPW}(z,g;\;x) = \frac{1}{n_{x}} \sum_{i:x_i=x} \Bigg\{\frac{I\{Z_i=z, G_i=g\}}{\hat{\psi}_i(z,g;x^{int},x^{out})}\cdot Y_i + \bigg(1-\frac{I\{Z_i=z,G_i=g\}}{\hat{\psi}_i(z,g;x^{int},x^{out})}\bigg)\cdot \hat{\mu}_i(z,g;x^{out})\Bigg\},
\end{equation}

\vspace{-1em}
\begin{equation}
\begin{split}
\hat{\mu}^{SAIPW}(z,g;\;x) &= \frac{1}{n_{x}} \sum_{i:x_i=x} \Bigg\{\bigg(\frac{1}{n}\sum_{k=1}^n \frac{I\{Z_k=z, G_k=g\}}{\hat{\psi}_k(z,g;x^{int},x^{out})}\bigg)^{-1}\frac{I\{Z_i=z, G_i=g\}}{\hat{\psi}_i(z,g;x^{int},x^{out})}\cdot Y_i \\
    &+ \bigg[1-\bigg(\frac{1}{n}\sum_{k=1}^n \frac{I\{Z_k=z, G_k=g\}}{\hat{\psi}_k(z,g;x^{int},x^{out})}\bigg)^{-1}\frac{I\{Z_i=z,G_i=g\}}{\hat{\psi}_i(z,g;x^{int},x^{out})}\bigg]\cdot \hat{\mu}_i(z,g;x^{int},x^{out})\Bigg\}.
\end{split}
\end{equation}

\noindent In the above definitions, $n_{x}$ refers to the number of outcome units such that $x_i=x$, and the propensity score estimates $\hat{\psi}_i(z,g;x^{int},x^{out})$ and outcome model predictions $\hat{\mu}_i(z,g;x^{int},x^{out})$ are as defined in Section~\ref{sec:pate}. In practice, the subgroups defined by $x$ can either be pre-specified a priori or chosen in a data-driven manner, as described below.

\subsection{Data-Driven Subgroup Discovery}
\label{sec:subgroupdiscovery}

In many application areas, including environmental health, there is limited literature to inform the a priori specification of subgroups that may demonstrate effects differing from the population average. To address this issue, numerous heterogeneous subgroup discovery methods have recently been proposed, which offer a data-driven approach to identify subgroups of the population (defined via covariate values) whose effects differ most from the population average \citep{Kennedy2020,bargaglistofficre}. Having developed heterogeneous treatment effect estimators for the BNI setting, we now describe how to integrate these estimators within existing subgroup discovery approaches to allow for the identification of heterogeneous subgroups without a priori specification.

Using direct and spillover effect estimates from Section~\ref{sec:cate} computed for each outcome unit's observed covariate values (henceforth referred to as individualized average treatment effect, or IATE, estimates), we employ a method adapted from the DR-Learner proposed by \cite{Kennedy2020}. We start by subtracting the average treatment effect from each IATE, and regress these new de-meaned IATEs on outcome-unit and key-associated intervention-unit covariates. Continuous covariates could be utilized here, though in practice we recommend categorizing these covariates first for enhanced interpretability. The resulting model coefficients are then taken to represent estimated additive deviations from the average treatment effect for the particular subgroup. We report the coefficient estimate and 95\% CI for each covariate to characterize heterogeneity in the estimated treatment effects and identify subgroups that benefit or malefit more from the intervention as compared to the overall population. Potential alternatives to the DR-Learner approach include tree-based subgroup discovery methodology as proposed in \cite{bargaglistofficre}.

\subsection{Bootstrap Variance Estimation}\label{sec:boot}

In order to obtain uncertainty estimates for the proposed estimators in Sections~\ref{sec:pate} and \ref{sec:cate}, we implement a modified bootstrap procedure by repeating the following steps a large number of times: 
\begin{enumerate}
    \item Resample $n$ outcome units with replacement.
    \item Obtain the subset of intervention units that serve as key-associated and/or upwind units for at least one of the outcome units in the resampled outcome data set, keeping intervention-level covariates $\eta(\bm{x}^{out}_{1:n})$ fixed.
    \item Model and recompute propensity scores with the new subset of intervention units.
    \item Truncate propensity scores as needed and map propensity scores back to the outcome level.
    \item Compute both population average and heterogeneous treatment effect estimates.
\end{enumerate}
Then, using the collection of bootstrapped estimates of the treatment effects, we use the percentile method to obtain 95\% confidence intervals for each one.

Within the context of our motivating example, this bootstrap approach corresponds to a scenario in which the probability of intervention on a power plant is affected by all ZIP-level covariates from all ZIP codes, rather than just the resampled ZIP codes (hence, keeping intervention-level covariates fixed), but not all ZIP codes are included in the analysis. This is opposed to an alternative approach in which intervention-level covariates are recomputed based on only the set of remaining outcome units in each resampled data set, which would equate to the non-resampled outcome units never having existed. This approach allows for sufficient sampling variability without the use of other methods such as cluster bootstrap, in which arbitrary clusters would need to be induced.

\section{Simulations}
\label{sec:simulations}

\subsection{Simulation Approach}

In this section, we outline a novel simulation approach for the BNI setting, with which we illustrate the application of our proposed estimators and evaluate their performance. Particularly for BNI, it is difficult to generate synthetic interference structures that mimic realistic network interactions. To our knowledge, simulation designs in the literature for BNI settings have not previously considered treatment effect heterogeneity. 

Here, we propose a simulation approach that combines real-world interference structures and covariate data with simulated treatments and outcomes in order to maintain flexibility. The general approach is outlined in Figure \ref{fig:simapproach}, with details provided in Algorithm \ref{alg:simapproachalg}. The aim of this proposed simulation structure is to provide a roadmap for reconciling the complexity created by having two distinct intervention and outcome unit levels in the bipartite setting, while allowing for flexibility in the choices of mapping functions and summary measures.%, depending on context and any subject matter knowledge.

\spacingset{1}

\begin{algorithm}[h!]

\caption{Simulation Approach}\label{alg:simapproachalg} \small
\begin{enumerate}
    \item Based on the real-world interference structure, determine the appropriate key-associated and upwind treatment units for each outcome unit.
    \vspace{-2mm}
    \begin{enumerate}[(i)]
        \item If necessary, map key-associated and upwind treatments to binary values such that each outcome unit is associated with some binary treatment vector $T \in \{0,1\}^2$.
    \end{enumerate}
    \item Map outcome-level covariates to the intervention unit level:
    \vspace{-2mm}
    \begin{enumerate}[(i)]
        \item For each intervention unit, determine outcome units for which the given intervention unit is key-associated and upwind.
        \item Create a summary measure of the outcome-level covariates that each intervention unit is key-associated or upwind for.
    \end{enumerate}
    \item Generate treatments at the intervention unit level:
    \vspace{-2mm}
    \begin{enumerate}[(i)]
        \item Using outcome-level covariate summaries and intervention-level covariates, assign treatments according to the desired mechanism.
        \item Compute propensity scores for each intervention unit.
    \end{enumerate}
    \item Map treatment assignments and propensity scores to the outcome-level based on key-associated and upwind treatments:
    \vspace{-2mm}
    \begin{enumerate}[(i)]
        \item Truncate extreme values of propensity scores, as needed.
        \item Compute the joint propensity score using the individual key-associated and upwind components.
    \end{enumerate}
    \item Repeat Monte Carlo iterations:
    \vspace{-2mm}
    \begin{enumerate}[(i)]
        \item Generate draws of the outcome from the desired probability distribution.
        \item Compute estimated heterogeneous treatment effect contrasts within each a priori-identified subgroup.
        \item Compute desired evaluation metrics.
    \end{enumerate}
\end{enumerate}
\vspace{-2mm}
\end{algorithm}

\spacingset{2}

While this approach can be more generally applied depending on any particular application of interest, we invoke it to simulate data mimicking those in our motivating example of air pollution transport, scrubber installation, and IHD hospitalization rates. Our outcome and intervention units are the ZIP codes and power plants in the real data, respectively, and we utilize the real HyADS matrix to characterize the interference structure and the real ZIP code and power plant features as covariates. Synthetic treatment statuses are generated for each power plant. Then, in each Monte Carlo iteration, new synthetic potential outcomes are generated and the methods described in Section~\ref{sec:methods} are applied to the synthetic data. 

To determine the key-associated and upwind power plants for each ZIP code (step 1 in Algorithm~\ref{alg:simapproachalg}), we use the real HyADS matrix. For the purposes of this demonstration, we consider a simple mapping for the upwind treatment vector (both in the data generation and in the model implementation).
%, though in practice an investigator could choose any appropriate mapping of the key-associated and upwind treatments. 
For each ZIP code, we define the binarized upwind treatment to be the treatment status of the power plant with the second-highest HyADS score for that unit. Thus, the direct and upwind treatments are subsequently referred to as ``key-associated'' and ``second key-associated'' treatments respectively, corresponding to the treatment statuses of the two most highly-influential power plants for each ZIP code.

In step 2 of Algorithm~\ref{alg:simapproachalg}, for each power plant, we compile ZIP codes for which the plant is 1) key-associated, and 2) second key-associated, and obtain averages (corresponding to the $\eta$ function introduced in Section~\ref{sec:effects}) of each covariate as ``intervention-level summary covariates''. In this context, for the ZIP codes a power plant is associated with, these are summary measures of their outcome-level covariates (demographic, meteorological, etc.). For this study, ZIP codes with key-associated HyADS scores in the bottom 25$^{th}$ percentile were removed from the analysis, in order to avoid potentially unstable weights interfering with the analysis. The remaining data in this simulation study consists of 21,338 ZIP codes, for which 260 unique power plants are key-associated and 268 unique power plants are second key-associated based on their HyADS scores (note that power plants could be key-associated for certain ZIP codes and second key-associated for others).

Explicit details on treatment and outcome-generating processes are provided in Appendix~\ref{sec:simdetails} and described briefly here. Treatments $T_j$ are generated for each individual power plant $j$ (step 3 of Algorithm~\ref{alg:simapproachalg}), and intervention-level propensity scores $\phi(t; x^{int}, \bm{x}^{out})$ are estimated from a logistic regression model. These are then mapped to the outcome unit level, so that each outcome unit's $Z$ and $G$ values correspond to the treatment statuses of its key- and second-key associated power plants (step 4). Due to the multiplicative nature of the joint propensity score, we inevitably encounter a larger proportion of extreme propensity score values compared to the non-bipartite setting. Thus, we truncate the top and bottom 5\% of propensity scores in both $Z$ and $G$. In this example, the synthetically generated potential outcomes represent an arbitrary pollution-related health outcome and are functions of ZIP code covariates and associated treatment statuses (step 5). When generating the potential outcomes, we impose heterogeneity in both the direct and spillover effects, where subgroups with particular covariate combinations are assigned differing effect sizes.

%Simulations are conducted according to a number of different scenario specifications.
In our first simulation study, we seek to empirically assess the doubly-robust properties of the AIPW estimators. In the data generating mechanism, we fix both the population average direct and spillover effects, here defined as the true average direct and spillover effect experienced over the entire population and henceforth referred to as the PATE $\xi$, as well as the outcome model error variance to be 1. To each simulated dataset, we apply estimators where both the propensity score and outcome models are correctly specified and compare with estimators where one or both of these models are intentionally misspecified. The different misspecification scenarios are as follows:
\begin{itemize}
    \item \textbf{Scenario A:} Both the propensity score and outcome models are correctly specified.
    \item \textbf{Scenario B:} The propensity score model is misspecified, but the outcome model is correctly specified.
    \item \textbf{Scenario C:} The propensity score model is correctly specified, but the outcome model is misspecified.
    \item \textbf{Scenario D:} Both the propensity score and outcome models are misspecified.
\end{itemize}
Details on misspecified model specifications are provided in Appendix~\ref{sec:simdetails}.

In each of the following simulation studies, we correctly specify both the propensity score and outcome models and vary features of the data generating mechanism to assess performance under a variety of data structures. In the second simulation study, we consider the impacts of using smaller subsets of the population, rather than conducting estimation on the entire superpopulation itself. Using sample proportion sizes of $0.5,0.2,0.1,0.05,0.03,0.01,$ and $0.005$, we randomly sample from the population and compute estimator evaluation metrics over 1,000 iterations for each sample. In the third simulation study, we evaluate the effects of modifying the outcome model error variance. We conduct simulations using outcome model error variance values of $0.2, 1, 5,$ and $10$, maintaining $\xi=1$.
Lastly, in order to assess the effects of changing the PATE relative to the outcome model error variance, we modify $\xi = \{1,5,10\}$ while holding the error variance constant at $1$.

In each of the simulation scenarios described above, we generate 1,000 simulated datasets and apply the AIPW, SAIPW, and G-computation estimators to estimate the heterogeneous treatment effects for each subgroup (assuming subgroups with heterogeneous effects are known a priori). We then compute a modified version of percent absolute bias (hereafter, `AB') in the heterogeneous treatment effect estimates from each dataset, which we define as:
\[AB(\mathscr{S}) = \frac{\lvert TE_{est}(\mathscr{S}) - TE_{true}(\mathscr{S}) \rvert}{\xi} \times 100\]
where $TE_{est}(\mathscr{S})$ denotes the estimated direct or spillover effect and $TE_{true}(\mathscr{S})$ denotes the corresponding true direct or spillover effect for subgroup $\mathscr{S}$. The PATE $\xi$ is used rather than $TE_{true}(\mathscr{S})$ in the denominator of the expression above in order to avoid potential errors when $TE_{true}(\mathscr{S}) = 0$.

\subsection{Simulation Results}

We present the results of our simulation studies for each scenario described above. Some of these results are provided here; additional figures can be found in Appendix~\ref{sec:additionalfigures}. 

Figures \ref{fig:boxplots_scens_DE} and \ref{fig:boxplots_scens_SE} show boxplots of the AB in the direct and spillover effects, respectively, computed for each estimator over 1,000 replications in simulation study 1 (model misspecification). Results from the G-computation estimator are shown in red, with the proposed AIPW and SAIPW estimators shown in green and blue, respectively. In both the direct and spillover effects, we see significantly better performance from the AIPW and SAIPW estimators compared to the G-computation estimator when both the propensity score model and outcome model are correctly specified or only one of the two is misspecified. Only when both propensity score and outcome models are misspecified we see a relative decrease in performance of these estimators. 

\spacingset{1.5}
\begin{figure}[ht!]
    \centering
    \includegraphics[width=\linewidth]{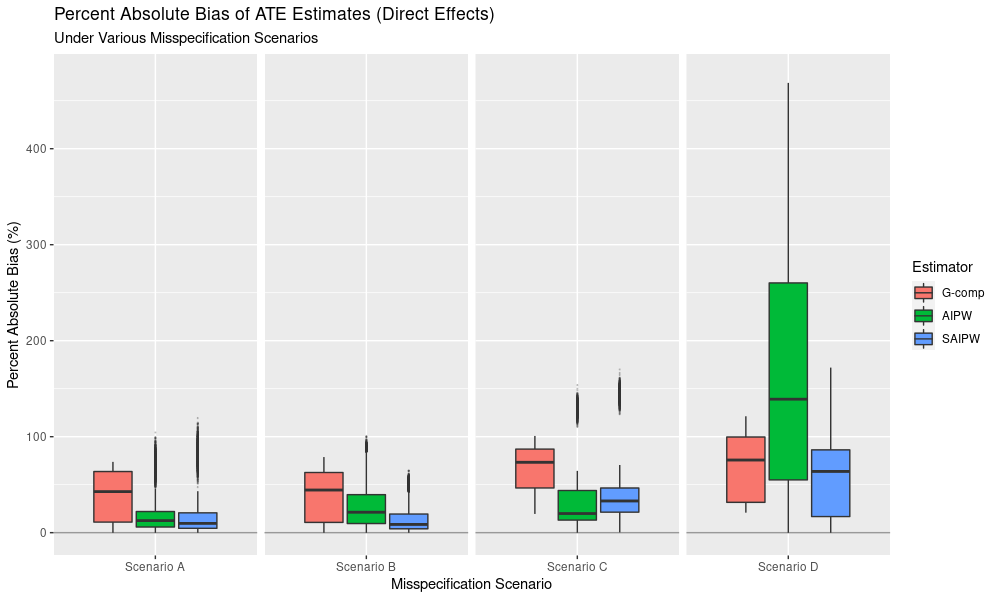}
    \caption{Percent absolute bias of subgroup average treatment effect estimates for direct effects, aggregated over all subgroups. Scenario specifications are detailed in Section~\ref{sec:simulations}.}
    \label{fig:boxplots_scens_DE}
\end{figure}

\begin{figure}[ht!]
    \centering
    \includegraphics[width=\linewidth]{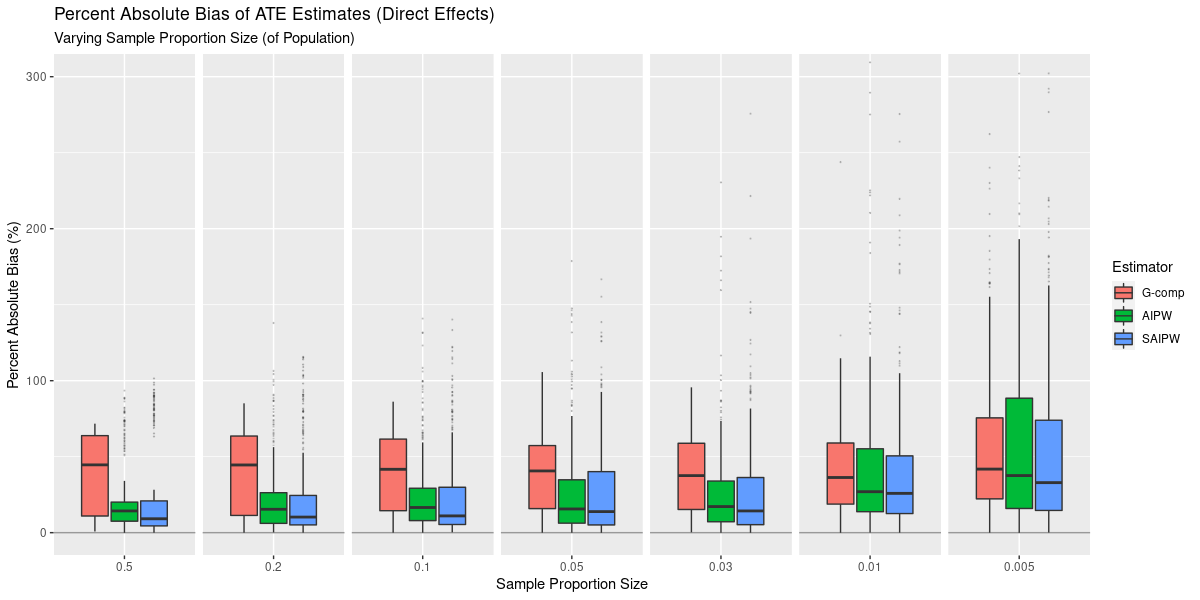}
    \caption{Percent absolute bias of subgroup average treatment effect estimates for direct effects over various sample proportions from the superpopulation, aggregated over all subgroups. This scenario uses correctly specified propensity score and outcome models.}
    \label{fig:samplesize_boxplots_DE}
\end{figure}

\spacingset{2}
AB for direct and spillover effects from simulation study 2 (varying sample sizes) are shown in Figures \ref{fig:samplesize_boxplots_DE} and \ref{fig:samplesize_boxplots_SE}, respectively. Even when decreasing the proportion of the population that is sampled, the AIPW and SAIPW estimators significantly outperform the G-computation estimator and are relatively unbiased for reasonably large sample sizes. The AB of the AIPW and SAIPW estimators tends to increase and converge to the AB of the G-computation estimator only around the point when the sample proportion decreases to around 0.01, corresponding to a sample size of about 200.

\spacingset{1.5}

\spacingset{2}
Figures \ref{fig:boxplots_vars_DE} and \ref{fig:boxplots_vars_SE} depict results under different outcome model error variance values, holding the PATE $\xi=1$. The AIPW and SAIPW estimators significantly outperform G-computation at when the variance is set to a value similar to the PATE. As the variance becomes extremely large relative to the PATE, making estimation more difficult, performance across the three estimators begins to even out.

Lastly, Figures \ref{fig:boxplots_eff_DE} and \ref{fig:boxplots_eff_SE} show results for different values of $\xi$, maintaining a constant outcome model error variance of 1. Across the three $\xi$ values, results remain consistent, with the AIPW and SAIPW estimators continuing to outperform G-computation in each case. Estimation improves as the magnitude of the PATE increases, which is expected as we hold the outcome model error variance constant.

\section{Application}
\label{sec:application}

\subsection{Details}
\begin{figure}
    \centering
    \includegraphics[scale=0.131]{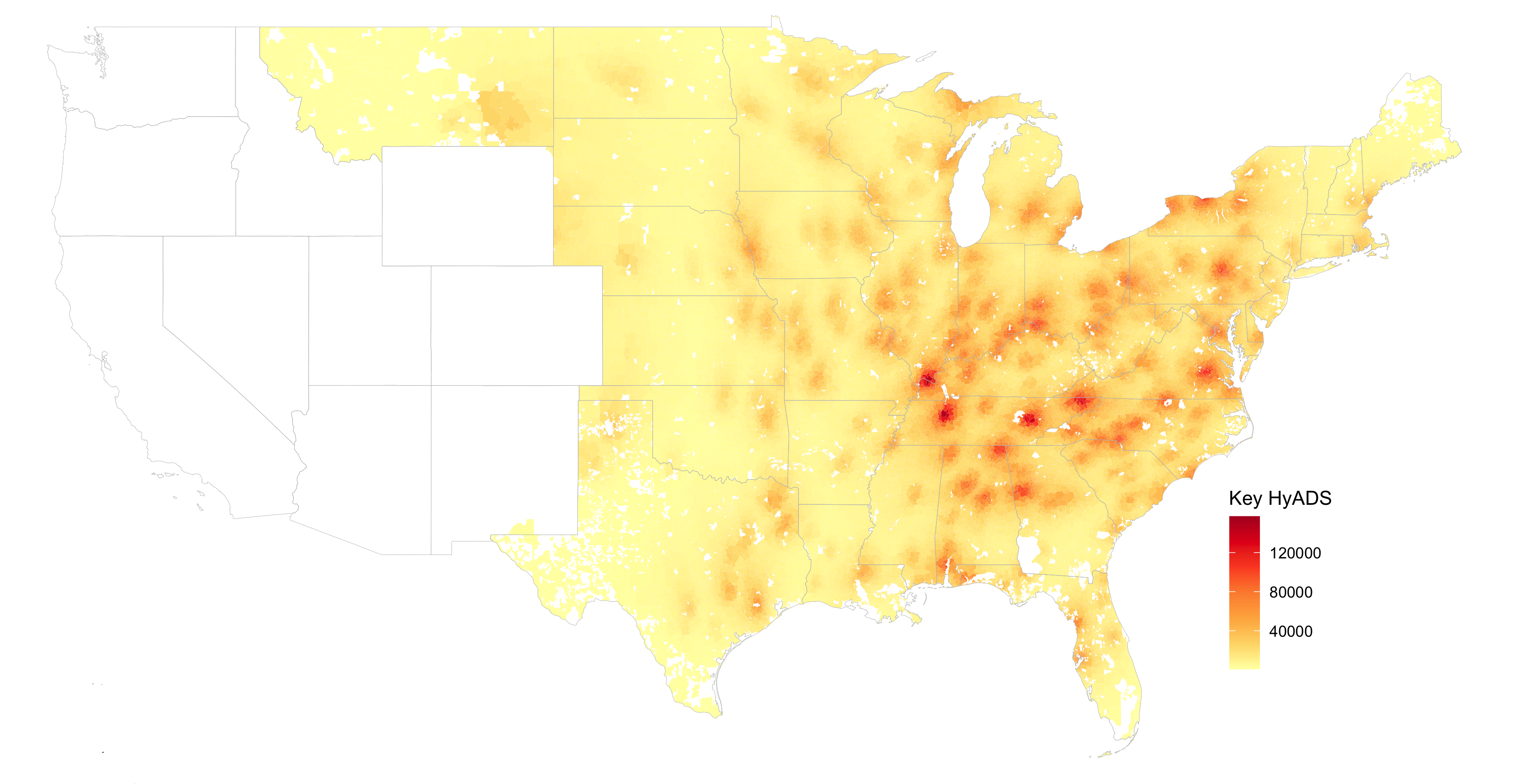}
    \caption{Maximum HyADS score experienced by each of the 29,304 U.S. ZIP codes in our study sample, used to determine each ZIP code's respective key-associated power plant ($j^*_i$).}
    \label{fig:hyads_map}
\end{figure}

We apply our proposed methods to study the impacts of FGD scrubber installations on power plants on IHD hospitalizations from 27,312,190 Medicare fee-for-service beneficiaries in the year 2005 (aggregated to create ZIP code level rates). Due to the relative lack of power plants in the Western US, we remove ZIP codes in the U.S. states of Wyoming, Colorado, Idaho, Utah, Arizona, New Mexico, Nevada, California, Oregon, and Washington. Due to the presence of extreme outlying outcomes in the upper tail (likely attributable to small population sizes creating extreme rates), we also excluded ZIP codes with IHD hospitalization rates in top 2\% in order to mitigate the potential for undue influence. This results in an overall sample of $n=29,304$ ZIP codes retained for analysis. A map of these ZIP codes and their key-associated HyADS values is provided in Figure \ref{fig:hyads_map}.

For this application, we adopt a mapping strategy for upwind treatment statuses similar to that of Section~\ref{sec:simulations}, in which we define the upwind treatment for each outcome unit as the treatment status of the power plant with the second-highest HyADS score associated with that unit. We estimate the propensity scores using a random forest model with both power plant covariates and summarized ZIP code covariates (as outlined in the simulation schema in Section~\ref{sec:simulations}). Power plant covariates and their descriptions are listed in Table \ref{tab:covariates}. We define $\eta(\cdot)$ (Section~\ref{sec:effects}) to be the simple mean function; thus, at each power plant, we obtain means of covariates for outcome units that are key-associated or second key-associated with that plant. To moderate the potential for extreme propensity scores resulting from the construction of the joint propensity scores, we truncate both the intervention-level and joint propensity scores. %In sensitivity analyses, we find that the particular level of truncation was not of significant consequence. 

\spacingset{1.5}

\spacingset{2}
Outcomes are defined as the IHD hospitalization rate among Medicare fee-for-service beneficiaries within each ZIP code in the year 2005. To specify the outcome model, IHD hospitalization rates are regressed on the outcome-level covariates listed in Table \ref{tab:covariates}. Estimates of direct effects (holding upwind treatments $G$ constant) and spillover effects (holding key-associated treatments $Z$ constant) are computed using the G-computation, AIPW, and SAIPW estimators detailed in Section~\ref{sec:methods}. 

\subsection{Results}

\subsubsection{Overall Population}
\label{sec:overallresults}
Treatment effect estimates for the overall population are summarized in Table \ref{tab:applicationresults}. 95\% bootstrap confidence intervals are computed as described in Section \ref{sec:boot} using 1,000 bootstrap samples. Estimates are reported as changes in the number of Medicare IHD hospitalizations per 10,000 person-years in 2005 as a result of FGD scrubber installations. Of particular note, in the overall sample population, we find that $\hat{\tau}(1) = -4.6 \; (\text{95\% CI:} [-27.4, 23.0])$ using the SAIPW estimator, meaning that installation of a scrubber at a ZIP code's most highly-associated power plant results in an average decrease of 4.6 IHD hospitalizations per 10,000 person-years, when that ZIP code's second-most influential power plant also has a scrubber installed. When the second-most influential power plant does not have a scrubber installed, the effect is estimated as $\hat{\tau}(0) = -22.1 \; (\text{95\% CI:} [-23.7, 0.2])$ hospitalizations per 10,000 person-years. However, using bootstrapped variance estimates, neither $\hat{\tau}(1)$ nor $\hat{\tau}(0)$ in the overall sample are significantly different from zero at the $\alpha=0.05$ level.

\spacingset{1.5}
\begin{table}[ht!]
\begin{adjustbox}{max width=\textwidth}
\begin{tabular}{@{}lccccc@{}}

 & $\hat{\tau}(G=1)$ \hspace{2mm} & $\hat{\tau}(G=0)$ \hspace{2mm} & $\hat{\delta}(Z=1)$ \hspace{2mm} & $\hat{\delta}(Z=0)$ \\
\midrule
Overall & & mean (95\% CI)\\
\hspace{16mm} G-comp & 2.7 (-12.8, 21.3) & -10.2 (-16.3, -4.2) & 6.6 (-10.4, 25.5) & -6.3 (-12.7, 0.6)\\
\hspace{16mm} AIPW & -8.6 (-30.6, 22.8) & -26.8 (-41.0, 14.7) & 9.5 (-38.7, 38.2) & -8.7 (-20.3, 1.2)\\
\hspace{16mm} SAIPW & -4.6 (-27.4, 23.0) & -22.1 (-23.7, 0.2) & 8.7 (-26.0, 25.8) & -8.9 (-21.2, 1.4)\\
% \cdashlinelr{1-5}
% \hspace{4mm} PctPoor $>$ Median(PctPoor)\\
% \hspace{16mm} G-comp & 1.5 (-23.3, 26.7) & -17.6 (-25.3, -10.1) & 8.0 (-16.1, 31.7) & -11.1 (-21.0, -2.2)\\
% \hspace{16mm} AIPW & -13.9(-47.7, 28.3) & -38.1 (-78.7, 9.2) & 11.1 (-44.0, 67.9) & -9.3 (-29.3, 3.5)\\
% \hspace{16mm} SAIPW & -9.3 (-42.9, 28.0) & -32.7 (-43.1, -6.4) & 10.2 (-34.0, 40.8) & -13.1 (-29.4, 3.7)\\
% \hspace{4mm} PctPoor $<$ Median(PctPoor)\\
% \hspace{16mm} G-comp & 3.9 (-13.7, 24.1) & -2.9 (-9.4, 4.1) & 5.3 (-12.2, 25.6) & -1.6 (-10.3, 6.9)\\
% \hspace{16mm} AIPW & -3.3 (-37.6, 35.7) & -15.4 (-23.1, 38.9) & 7.9 (-59.5, 32.8) & -4.3 (-22.8, 9.3)\\
% \hspace{16mm} SAIPW & 0.2 (-31.8, 36.6) & -11.5 (-11.6, 15.7) & 7.1 (-39.9, 26.9) & -4.6 (-22.9, 9.3)\\
% \cdashlinelr{1-5}
% \hspace{4mm} PctNonwhite $>$ Median(PctNonwhite)\\
% \hspace{16mm} G-comp & 4.2 (-15.0, 26.0) & -13.6 (-19.2, -7.8) & 9.9 (-9.0, 29.8) & -8.0 (-15.0, -0.5)\\
% \hspace{16mm} AIPW & -20.9 (-61.4, 29.9) & -39.0 (-84.0, -0.7) & 5.2 (-46.3, 71.1) & -12.9 (-30.3, 4.1)\\
% \hspace{16mm} SAIPW & -12.5 (-47.2, 31.2) & -32.1 (-41.6, -8.1) & 6.4 (-37.0, 41.0) & -13.2 (-31.3, 4.3)\\
% \hspace{4mm} PctNonwhite $<$ Median(PctNonwhite)\\
% \hspace{16mm} G-comp & 1.2 (-20.4, 28.4) & -6.9 (-14.9, 0.9) & 3.4 (-17.0, 30.3) & -4.7 (-13.6, 3.6)\\
% \hspace{16mm} AIPW & 3.7 (-23.5, 35.7) & -14.5 (-19.0, 46.7) & 13.8 (-55.4, 32.8) & -4.5 (-19.6, 8.1)\\
% \hspace{16mm} SAIPW & 3.3 (-23.6, 36.0) & -12.1 (-14.2, 14.3) & 10.9 (-28.9, 31.7) & -4.5 (-19.8, 8.8)\\
\bottomrule
\end{tabular}
\caption{Estimated change in Medicare IHD hospitalizations per 10,000 person-years in 2005 resulting from FGD scrubber installation at power plants with 95\% bootstrap confidence interval approximations.}
\label{tab:applicationresults}
\end{adjustbox}
\end{table}

\spacingset{2}
\subsubsection{Treatment Effect Heterogeneity}
\label{sec:hteresults}

To investigate potential effect heterogeneity, we implement the subgroup discovery approach outlined in Section~\ref{sec:subgroupdiscovery}, in which we de-mean and regress the treatment effects computed for each ZIP code in Section~\ref{sec:overallresults} on their ZIP-level covariates and key-associated power plant covariates. These covariates include demographic and socioeconomic variables measured in the Medicare population (mean age, female rate, and non-white rate) and in the population as a whole (population size, urbanicity, high school graduation rate, occupied housing rate, proportion of households moved within the previous 5 years, and smoking rate), weather (mean temperature and relative humidity), and power plant operating characteristics (total NO$_\text{x}$ controls, heat input, operating time, operating capacity, selective non-catalytic reduction rate, and whether the plant was in Phase 2 of the EPA's Acid Rain Program (ARP)). The choice of these variables was motivated by recent literature that found disparities in air pollution exposure and health effects between demographic and socioeconomic status groups, as well as the EPA's charge to further examine such inequities in environmental exposures and risks \citep{sacks2011particulate,jbaily2022air,Josey2023,usepa2022a}. Meteorology is also a known modifier of air pollution health impacts \citep{zanobetti2015disentangling,li2017modification,anenberg2020synergistic}. For ease of interpretation, each covariate was binarized according to its observed median value. To avoid undue influence of extreme values, an outlier-robust regression was used \citep{Venables2002}.

\begin{figure}[ht!]
    \includegraphics[width=\textwidth]{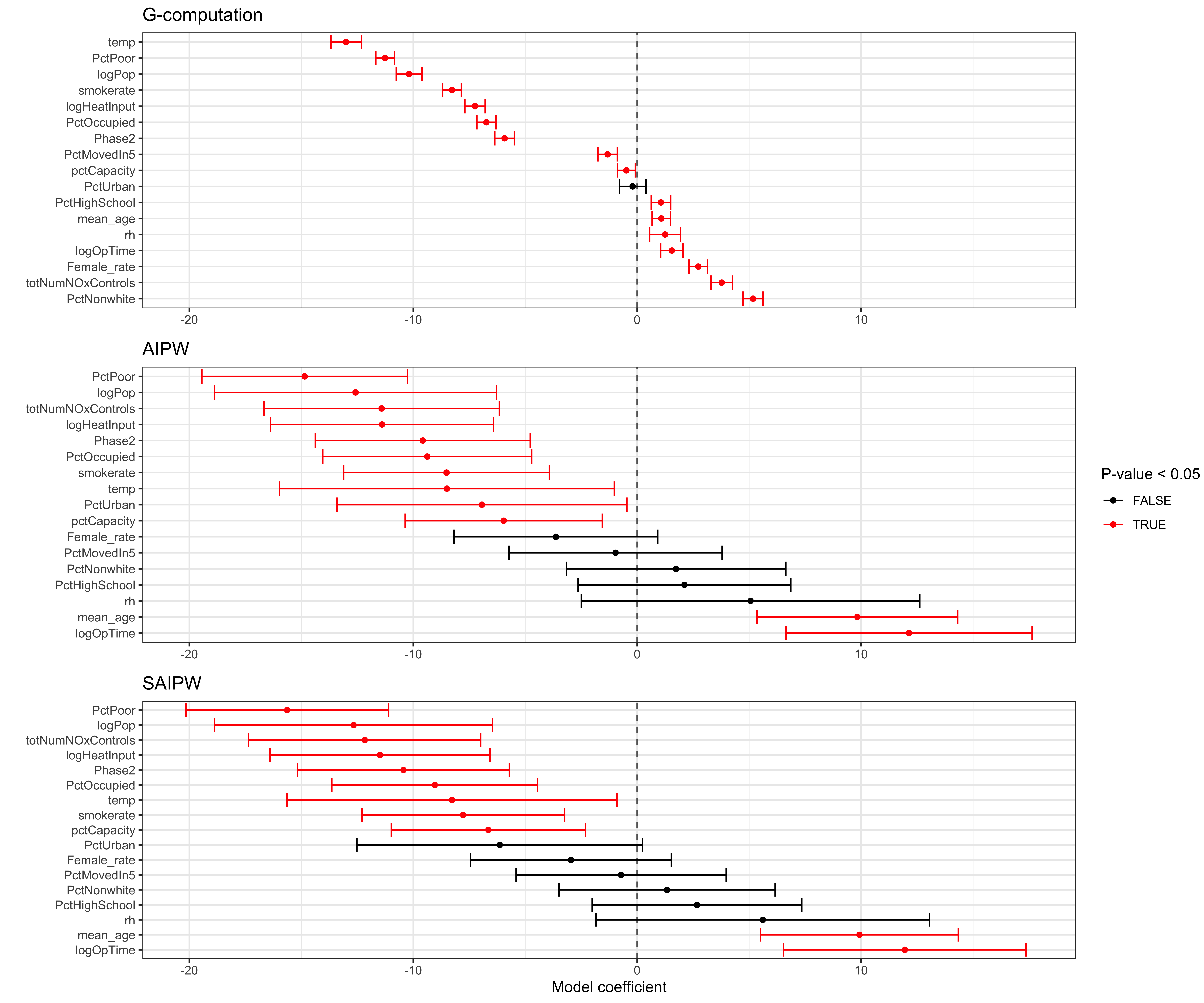}
    \caption{Model coefficients regressing de-meaned direct effects, holding upwind treatment at G=0, on individual ZIP covariates, representing heterogeneity of treatment effects by population subsets.}
    \label{fig:results_hte}
\end{figure}

For brevity, results of this analysis for $\hat{\tau}(G=0)$ are shown in Figure~\ref{fig:results_hte}. Additional results, including sensitivity analyses using various amounts of outlier trimming to demonstrate stability, are given in Appendix Figures~\ref{fig:de_g0_01_trim}-\ref{fig:se_z1_notrim}. Of particular interest, ZIP codes with higher-than-median poverty and smoking rates appear to receive significantly greater benefits from scrubber installation on their most highly-influential power plants compared to the overall population, when their upwind plants are untreated. In addition, areas with higher average temperatures also experienced greater protective health benefits from scrubber installation on their respective plants. Consistent positive health benefits are also observed for several key-associated power plant characteristics, including those with higher-than-median heat input, capacity, and those that were part of Phase II of the EPA Acid Rain Program. 

The evidence from this study suggests that, particularly when upwind power plants have not been intervened on, there is a positive health benefit in treating a region's most influential power plants, particularly for marginalized and minority groups. This demonstrates the importance of methodology in bipartite settings which enable the study of heterogeneity in treatment effects.

\spacingset{1.5}

\spacingset{2}

\section{Discussion}
\label{sec:discussion}

We have proposed heterogeneous treatment effect estimands and causal AIPW and SAIPW estimators of direct and spillover effects in the BNI setting. We have shown that these estimators can be used to estimate effects even when treatment effect heterogeneity is present in the population and also improve on G-estimation methods. They can also be used in combination with subgroup discovery methods to identify population subgroups that are most affected by treatment without needing to specify the subgroups a priori. We have also designed and implemented extensive simulation studies in the BNI setting. The simulation design combines the utility of real-world interference structures with the flexibility of allowing user-specified treatment and outcome generating mechanisms. This approach provides a path to reconciling the fact that two distinct unit dimensions (namely, the intervention and outcome units) exist in the BNI setting, in contrast to typical interference scenarios where treatments and outcomes are measured on the same units.
%Though our choices of simulation parameters and mapping functions were motivated by our particular application to power plant emissions and air pollution mitigation policy, our approach allows for generalization to a wider range of use cases in which subject matter knowledge can inform specific choices in the simulation process.
To our knowledge, our proposed approach is the first attempt to simulate data and scenarios with treatment effect heterogeneity in the BNI setting through careful consideration of each step involved in mapping between intervention and outcome units, while maintaining a tractable causal estimation problem. Using this simulation approach, we have shown empirically that our AIPW-based estimators are robust to misspecification in either the estimated propensity score or outcome models and that the performance of our methodology remains relatively strong even when applied to smaller samples.

Our methods are particularly relevant to air pollution regulatory strategies and policy development for a number of reasons. The study of heterogeneous treatment effects in the BNI setting is crucial for air pollution regulatory policy as it identifies subpopulations that may be most affected by policy decisions. Our findings show that, using the proposed methods, there was no statistically significant effect of scrubber installation on IHD hospitalization rates in the overall population. An investigation focused on only these population average effects would have likely concluded ``no effect'' and overlooked the health benefits of scrubbers for certain subgroups. In particular, we observe statistically significant health benefits in marginalized subgroups and low-income communities. Increased health benefits of emissions reductions for these population subgroups are scientifically plausible due to the presence of social and structural systems that lead to differing degrees of exposure and vulnerability \citep{Josey2023}. For example, low-income individuals are disproportionately likely to work outdoors and to lack access to air conditioning, thus increasing their exposure to ambient air pollution. This suggests that studies of population average treatment effects may not be sufficient to fully characterize the effects of environmental policies on health outcomes. Our results highlight the importance of methodology that accounts for treatment effect heterogeneity in order to design effective policies in an environmental justice context.

There are several limitations of our methods to note. In our simulations, simplifications were made to map between intervention and outcome units. One example of this is in the estimation of propensity scores on the intervention unit level, in which summary measures of outcome unit covariates were mapped to each intervention unit. These decisions were made to maintain model parsimony; however, in our general approach, we have allowed for user specification based on particular applications or needs. Additionally, we make an implicit assumption of independence between outcome units conditional on observed ZIP-level and power plant covariates during estimation. Another limitation is in the binarization of both direct and spillover effects, which was done in order to limit the number of potential outcomes needing consideration in the BNI setting. This results in considerable information loss and often requires arbitrary redefinitions of direct and indirect treatments. In the power plant emissions scenario, the framing of one power plant being most influential, with all others being defined in some binary mapping as ``upwind'', may not reflect the true pollution transport mechanisms at any given outcome unit. These limiting factors motivate a need for a more nuanced approach to our definition of exposures. One potential extension that mitigates this problem is to consider a continuous treatment summary measure instead (e.g., collapsing the treatment vector using an interference network-weighted mean). This is a subject of future work.

%We also highlight a few additional extensions that follow from this paper. In ongoing work, we utilize our heterogeneous treatment effect estimators in subgroup discovery methods such as the Causal Rule Ensemble (CRE) method \citep{bargaglistofficre} in order to not only estimate heterogeneous effects among subgroups but to also discover heterogeneous subgroups without a priori specification. An application of this method could, for example, yield information on marginalized communities that would experience effects significantly differing from the population average without prior speculation. %Ongoing work also includes extensions to studying optimal treatment regimes in the BNI setting, in order to maximize potential benefits from large-scale intervention schemes. For example, a potential application of this idea is the determination of which power plants to outfit with an emissions-reduction treatment to maximize health benefits, while considering complex interference structures as well as limiting factors such as cost.

\section*{Funding and Acknowledgements}

This work was funded by NIH grants T32ES007142, T32GM135117, and K01ES032458.

\vspace{1em} 

\noindent We benefited from helpful comments and suggestions from Francesca Dominici, Cory Zigler, Laura Forastiere, Nima Hejazi, Kosuke Imai, Fabrizia Mealli, Georgia Papadogeorgou, Luke Miratrix, and participants at the Harvard Causal Inference workshop, at the NSAPH Biostatistics seminar, and at the 2022 IMS International Conference on Statistics and Data Science (ICSDS).

\clearpage

\bibliographystyle{biom}
\bibliography{references}

\clearpage

%TC:ignore
\section*{Appendix}
\renewcommand{\thesubsection}{\Alph{subsection}}
\renewcommand\thefigure{\thesubsection.\arabic{figure}}   
\renewcommand\thetable{\thesubsection.\arabic{table}}   
\setcounter{page}{1}

\subsection{Assumptions}\label{sec:assumptions}

\begin{assumption}[Overlap]
    \[0 < \phi_j(t; \; x^{int}, \; \bm{x}^{out}) < 1 \quad \forall j \in \{1,\dots,J\}.\]
\end{assumption}
\noindent The overlap assumption states that there is a strictly positive probability of treatment at each intervention unit $j$.

\begin{assumption}[Consistency]
If $Z_i=z$ and $G_i=g$, then \[Y_i = Y_i(z,g).\]
\end{assumption}
\noindent This assumption states that an individual's potential outcome under their observed key-associated and upwind treatments is exactly their observed outcome.

In addition to the above, we must also make an analogous assumption to the unconfoundedness assumption in classical causal inference. With BNI, we have two levels of potential confounders–-those at the intervention unit level, and those at the outcome unit level. Allowing for both sets of confounders, we assume the following:
\begin{assumption}[Ignorability]
\[Y_i(z,g) \indep Z_i,G_i \mid \bm{X}_i^{out}, \{\bm{X}_j^{int}\}_{j \in \{j_{(i)}^*,\bm{\underline j}_{(i)}^\dagger\}} \quad \forall i \in \{1,\dots,n\}, \forall z,g \in \{0,1\}.\]
\end{assumption}
\noindent The ignorability assumption states that a unit's potential outcomes are independent of its key-associated treatment and upwind treatment, conditioning on that outcome unit's covariates, its key-associated intervention unit's covariates, and its upwind intervention units' covariates.

\subsection{Simulation Details} \label{sec:simdetails}
\setcounter{figure}{0}

\begin{figure}[ht!]
    \centering
    \includegraphics[width=0.9\textwidth]{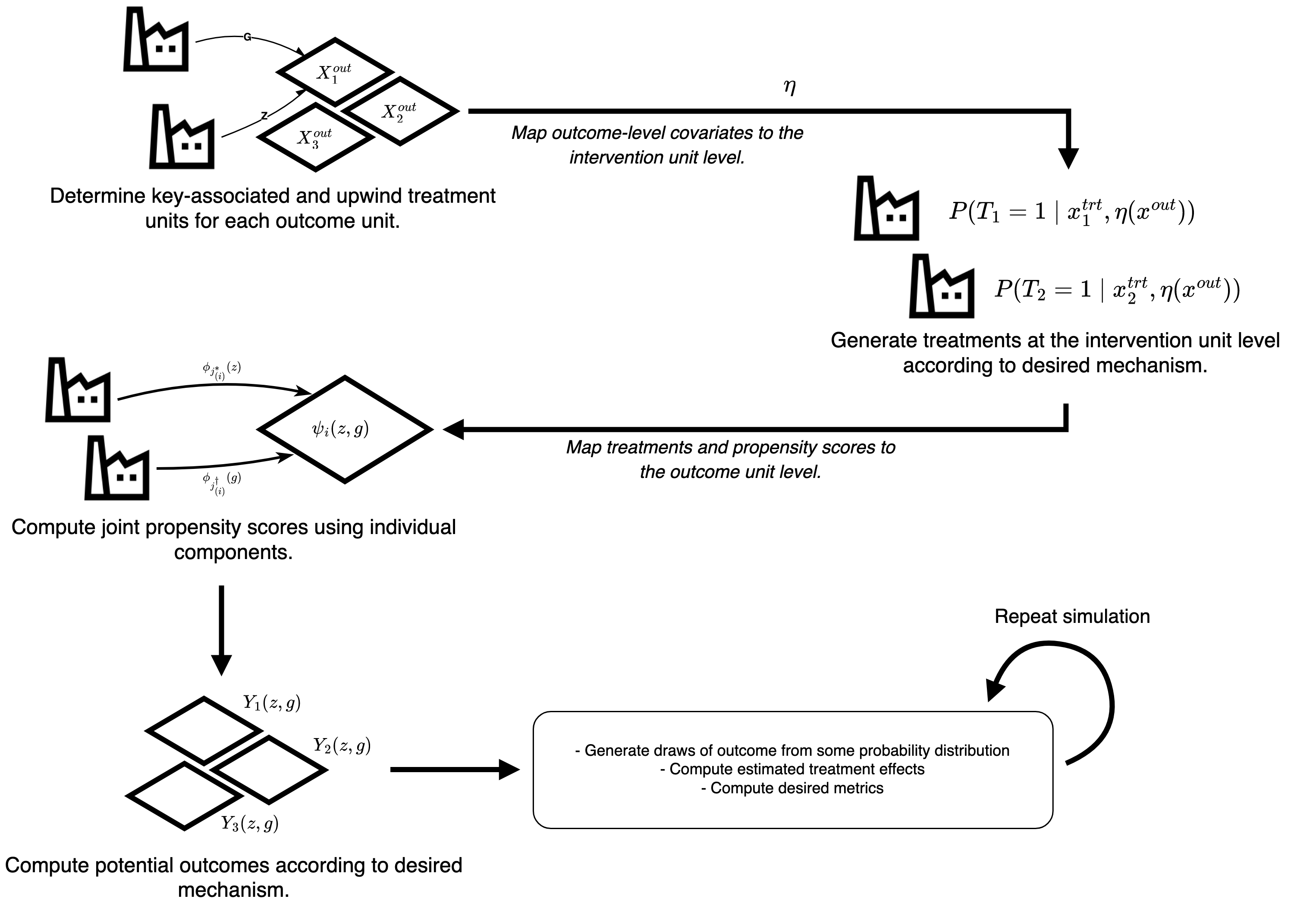}
    \caption{Flow chart depicting proposed BNI simulation approach.}
    \label{fig:simapproach}
\end{figure}

In our proposed simulation approach, we leverage the use of real-world ZIP code and power plant covariates in order to create realistic simulation scenarios motivated by their potential applications. At the intervention unit level, these covariates include those created by summarizing each power plant's key-associated and upwind ZIP code covariates in addition to its own plant characteristics, while on the outcome unit level, covariates are comprised of each ZIP code's own characteristics. Out of the available covariates, we consider a subset for the purposes of this simulation study, which we describe further in detail below, in addition to the generating model specifics used in our simulations. The general approach is outlined in Figure~\ref{fig:simapproach}.

\subsubsection*{Treatment Generating Models}

We include the following covariates at the intervention unit level: 
\begin{itemize}
    \item \(KeyLogPop_j\), the mean of the natural log population size in ZIP codes plant $j$ is key-associated for;
    \item\(KeyPctUrban_j\), the mean of the proportion of residents living in urban areas in ZIP codes plant $j$ is key-associated for;
    \item \(LogOpTime_j\), the log time of operation over one year, in hours, for plant $j$.
\end{itemize}

Treatments are assigned for intervention unit $j$ from a Bernoulli($p_j$) distribution, where \[\text{logit } p_j = 0.1KeyLogPop_j - 1.5KeyLogPop_j*KeyPctUrban_j + 0.05LogOpTime^2\]
This treatment assignment mechanism results in an overall proportion of treated plants of 0.462, which was chosen to promote balance in the treatment groups after mapping to the outcome level. After mapping, the following matrix shows the number of ZIP codes within each treatment group:

\begin{table}[ht!]
\begin{tabular}{l|ll|l}
        & $G_i=1$ & $G_i=0$ & Total \\ \hline
$Z_i=1$ & 4834    & 4286    & 9120  \\
$Z_i=0$ & 4104    & 8114    & 12218 \\ \hline
Total   & 8938    & 12400   & 21338
\end{tabular}
\end{table}

In misspecification scenarios B and D, in which treatment model is misspecified, we specify a logistic regression model with only linear terms to estimate propensity scores.

\subsubsection*{Outcome Generating Models}

For the outcome, we consider a subset of covariates on the outcome-level:
\begin{itemize}
    \item \(LogPop_i\), the log population size of ZIP code $i$;
    \item \(SmokeRate_i\), the proportion of smokers in ZIP code $i$;
    % \item \(MeanAge_i\), the mean age of the Medicare population in ZIP code $i$;
    \item \(PctHighSchool_i\), the proportion of high school graduates in ZIP code $i$;
    \item \(PctUrban_i\), the proportion of the population residing in an urban area in ZIP code $i$;
    \item \(PctPoor_i\), the proportion of the population living under the poverty threshold in ZIP code $i$;
    \item \(PctNonwhite_i\), the proportion of non-white residents in ZIP code $i$.
\end{itemize}

Heterogeneity is introduced in both the direct and spillover effects based on two outcome-level covariates, \(PctNonwhite_i\) and \(PctPoor_i\). Based on these covariates, we partition the population into three roughly equal-sized subgroups, on which we introduce heterogeneity for direct effect $\tau_i$ and spillover effect $\delta_i$, with PATE $\xi$:
\begin{itemize}
    \item $\tau_i = \delta_i = 2\xi$, if $PctNonwhite_i$ exceeds the 33$^{rd}$ percentile and $PctPoor_i$ exceeds the median value for all ZIP codes.
    \item $\tau_i = \delta_i = \xi$, if $PctNonwhite_i$ exceeds the 33$^{rd}$ percentile and $PctPoor_i$ lies below the median value for all ZIP codes.
    \item $\tau_i = \delta_i = 0$, if $PctNonwhite_i$ lies below the 33$^{rd}$ percentile of values for all ZIP codes.
\end{itemize}

Potential outcomes are assigned according to the following distribution:
\[Y_i(z,g) = N(\mu_i(z,g),\sigma^2)\]
where 
\begin{align*}
\mu_i(z,g) &= 2LogPop_i + 5SmokeRate_i + 5PctPoor_i + 10PctNonwhite_i\\
&+ 5PctNonwhite_i*SmokeRate_i + \tau_i z + \delta_i g
\end{align*}

In misspecification scenarios C and D, where the outcome model is misspecified, we specify a linear regression model without interaction terms.

\clearpage
\subsection{Additional Simulation Tables \& Figures} \label{sec:additionalfigures}
\setcounter{figure}{0}

\spacingset{1.5}
\begin{figure}[ht!]
    \centering
    \includegraphics[width=0.8\linewidth]{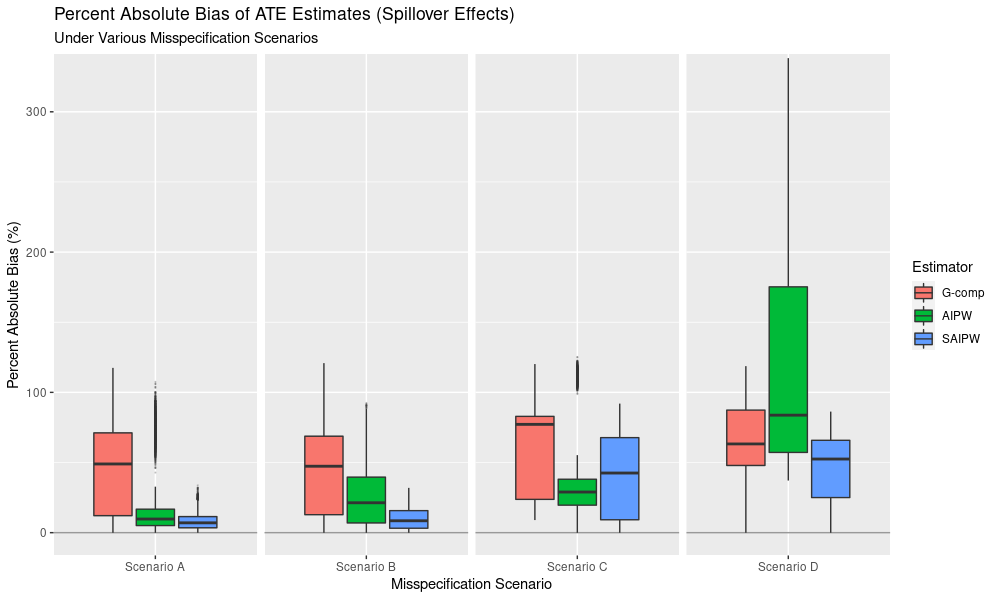}
    \caption{Percent absolute bias of subgroup average treatment effect estimates for spillover effects. Scenario specifications detailed in Section~\ref{sec:simulations}.}
    \label{fig:boxplots_scens_SE}
\end{figure}

\begin{figure}[ht!]
    \centering
    \includegraphics[width=0.9\linewidth]{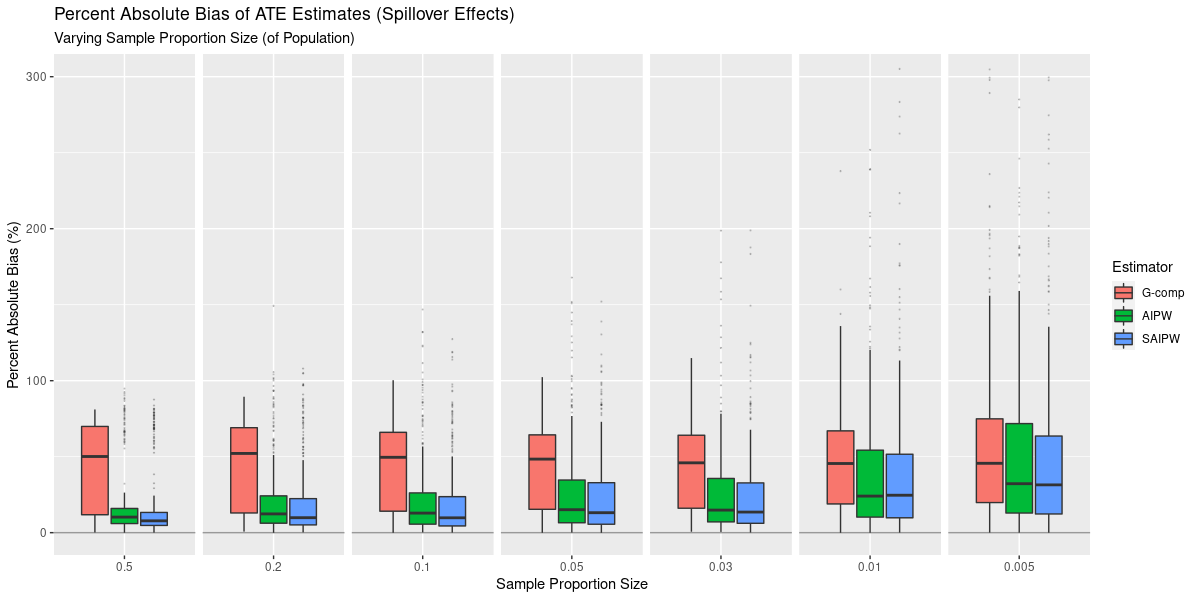}
    \caption{Percent absolute bias of subgroup average treatment effect estimates for spillover effects over various sample proportions from the superpopulation. Scenario uses correctly specified propensity score and outcome model.}
    \label{fig:samplesize_boxplots_SE}
\end{figure}

\begin{figure}[ht!]
    \centering
    \includegraphics[width=0.85\linewidth]{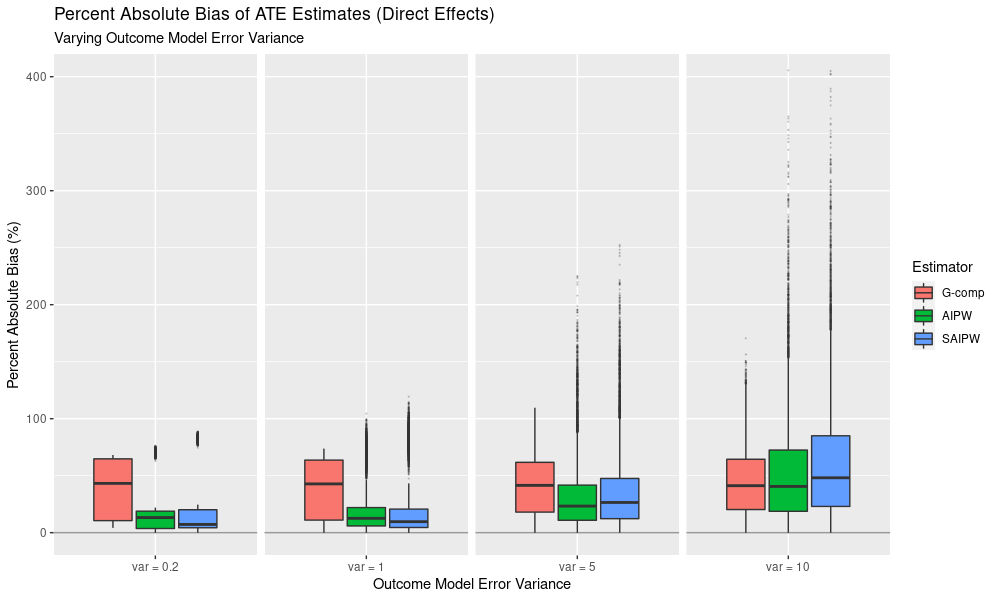}
    \caption{Percent absolute bias of subgroup average treatment effect estimates for direct effects over different outcome model error variances. Scenario uses correctly specified propensity score and outcome model.}
    \label{fig:boxplots_vars_DE}
\end{figure}

\begin{figure}[ht!]
    \centering
    \includegraphics[width=0.85\linewidth]{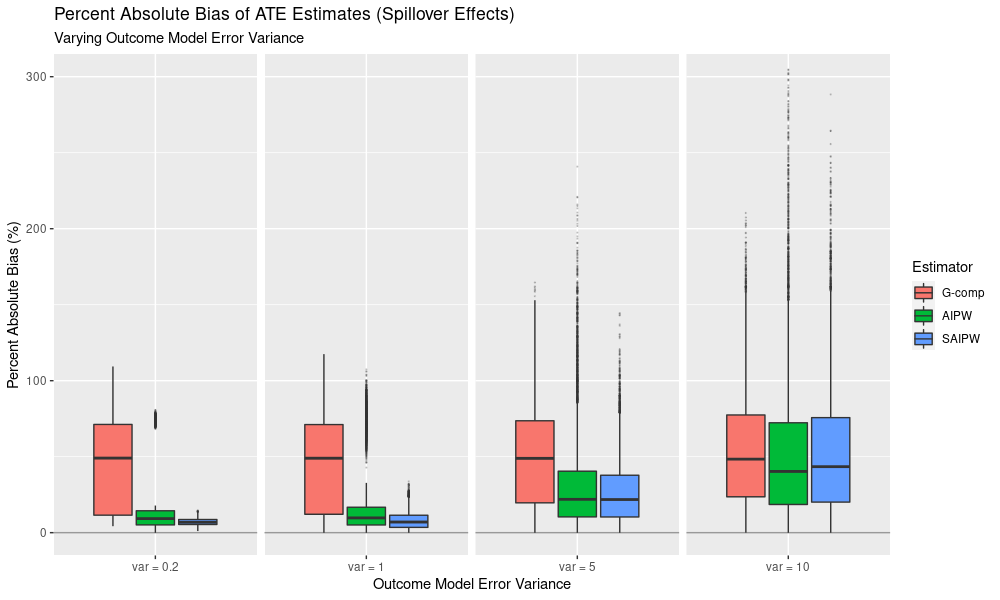}
    \caption{Percent absolute bias of subgroup average treatment effect estimates for spillover effects over different outcome model error variances. Scenario uses correctly specified propensity score and outcome model.}
    \label{fig:boxplots_vars_SE}
\end{figure}

\begin{figure}[ht!]
    \centering
    \includegraphics[width=0.85\linewidth]{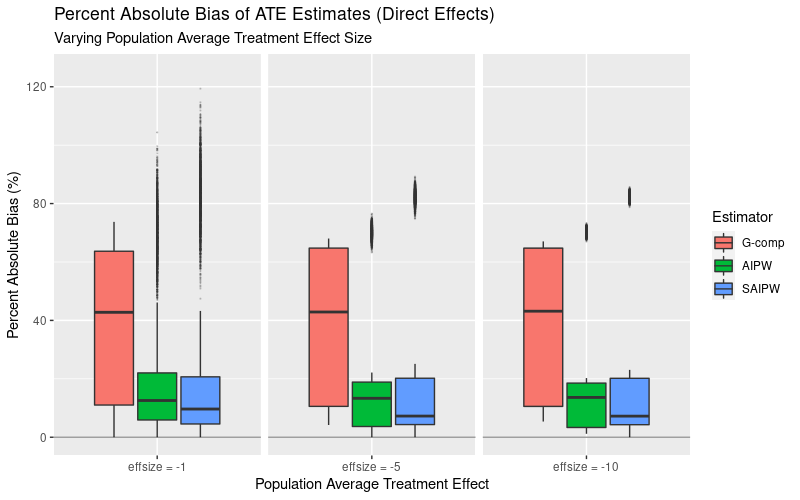}
    \caption{Percent absolute bias of subgroup average treatment effect estimates for direct effects over different population average effect sizes. Scenario uses correctly specified propensity score and outcome model.}
    \label{fig:boxplots_eff_DE}
\end{figure}

\begin{figure}[ht!]
    \centering
    \includegraphics[width=0.85\linewidth]{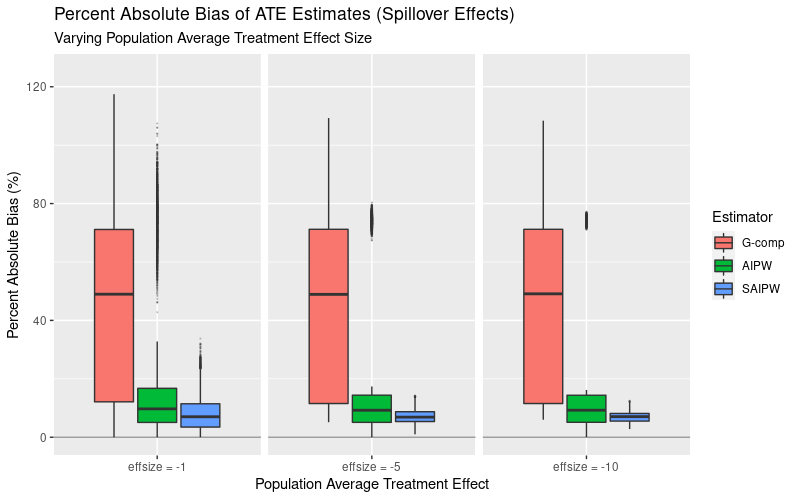}
    \caption{Percent absolute bias of subgroup average treatment effect estimates for spillover effects over different population average effect sizes. Scenario uses correctly specified propensity score and outcome model.}
    \label{fig:boxplots_eff_SE}
\end{figure}

\clearpage
\subsection{Application Covariates}
\setcounter{table}{0}

\begin{table}[ht!]
\begin{tabular}{llll}
Variable                            & Unit Level  & Mean   & Range          \\ \hline
log(Population)                     & ZIP         & 8.27  & (1.39, 11.65)  \\
\% Urban                             & ZIP         & 0.41  & [0, 1]         \\
\% High school graduate              & ZIP         & 0.35  & [0, 1]         \\
\% Poverty                           & ZIP         & 0.12  & [0, 1]         \\
\% Households occupied               & ZIP         & 0.88  & [0.01, 1]      \\
\% Moved in last 5 years             & ZIP         & 0.42  & [0, 1]         \\
Smoke rate                          & ZIP         & 0.26  & (0.10, 0.43)   \\
Mean temperature (K)                & ZIP         & 286.8 & (272.5, 299.1) \\
Mean relative humidity (\%)          & ZIP         & 0.008 & (0.003, 0.015) \\
Mean Medicare age                   & ZIP         & 74.90 & (68, 96.26)    \\
\% Female                            & ZIP         & 0.56  & [0, 1]         \\
\% Non-white                         & ZIP         & 0.11  & [0, 1]         \\ \hline
Total NO$_x$ controls               & Power plant & 4.41  & (0, 24)        \\
log(Heat input)                     & Power plant & 14.97 & (10.04, 17.32) \\
log(Operating time)                 & Power plant & 7.75  & (5.46, 8.93)   \\
\% Operating capacity                & Power plant & 0.63  & (0.07, 1.10)   \\
\% Selective non-catalytic reduction & Power plant & 0.18  & [0, 1]         \\
ARP Phase II                        & Power plant & 0.71  & \{0, 1\}  
\end{tabular}
\caption{Summary of covariates used in data application.}
\label{tab:covariates}
\end{table}

\clearpage
\subsection{Additional Application Results}
\setcounter{figure}{0}

Additional analysis results of treatment effect heterogeneity from Section~\ref{sec:hteresults} are provided below. These include sensitivity analyses after trimming the top and bottom 1\% and 5\% of outcomes and additional results for $\hat{\tau}(G=1)$, $\hat{\delta}(Z=1)$, and $\hat{\delta}(Z=0)$.

\begin{figure}[ht!]
    \centering
    \includegraphics[width=\linewidth]{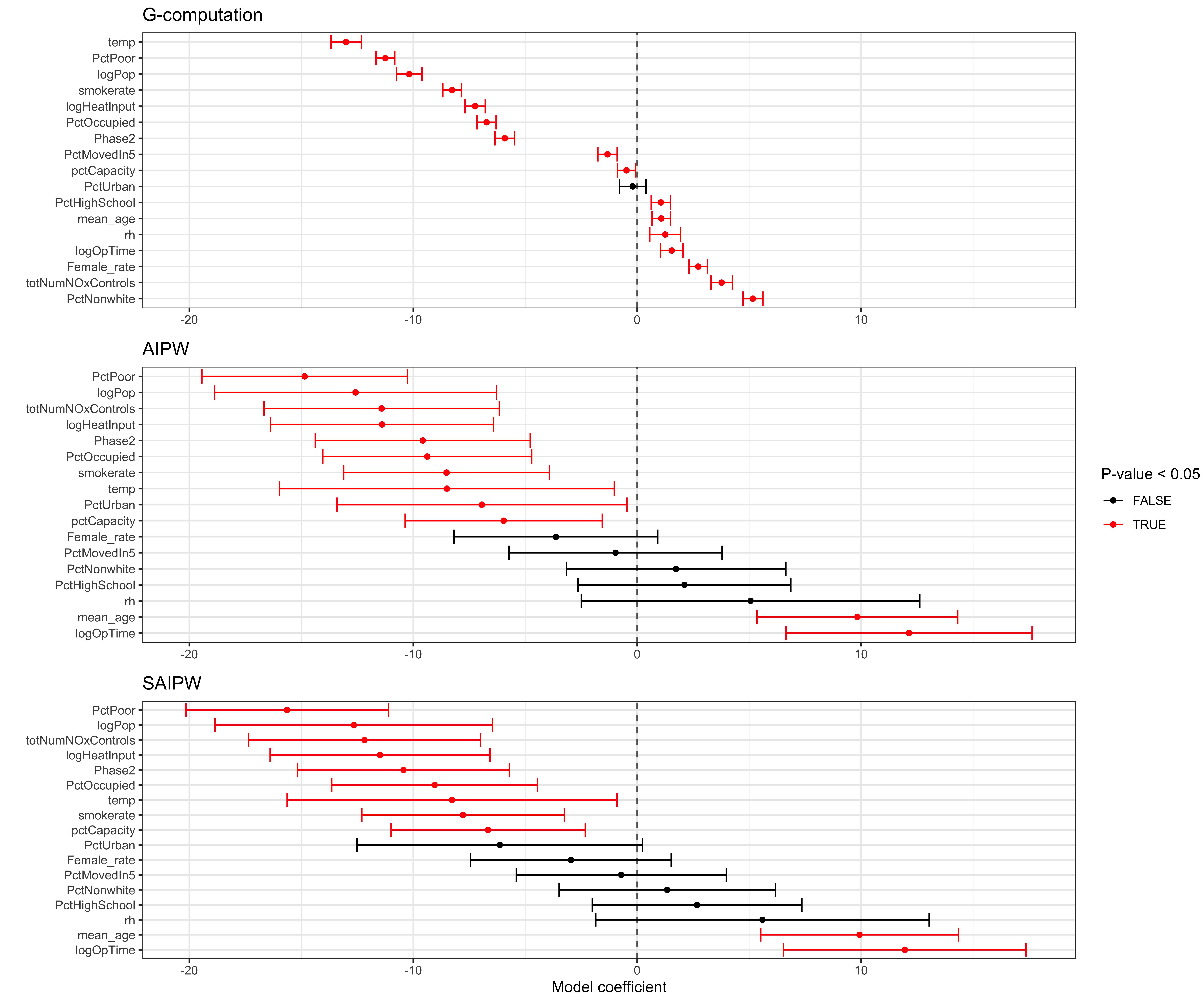}
    \caption{Heterogeneity of direct effects, holding upwind treatment at G=0, with the top and bottom 1\% of outcomes trimmed.}
    \label{fig:de_g0_01_trim}
\end{figure}

\begin{figure}[ht!]
    \centering
    \includegraphics[width=\linewidth]{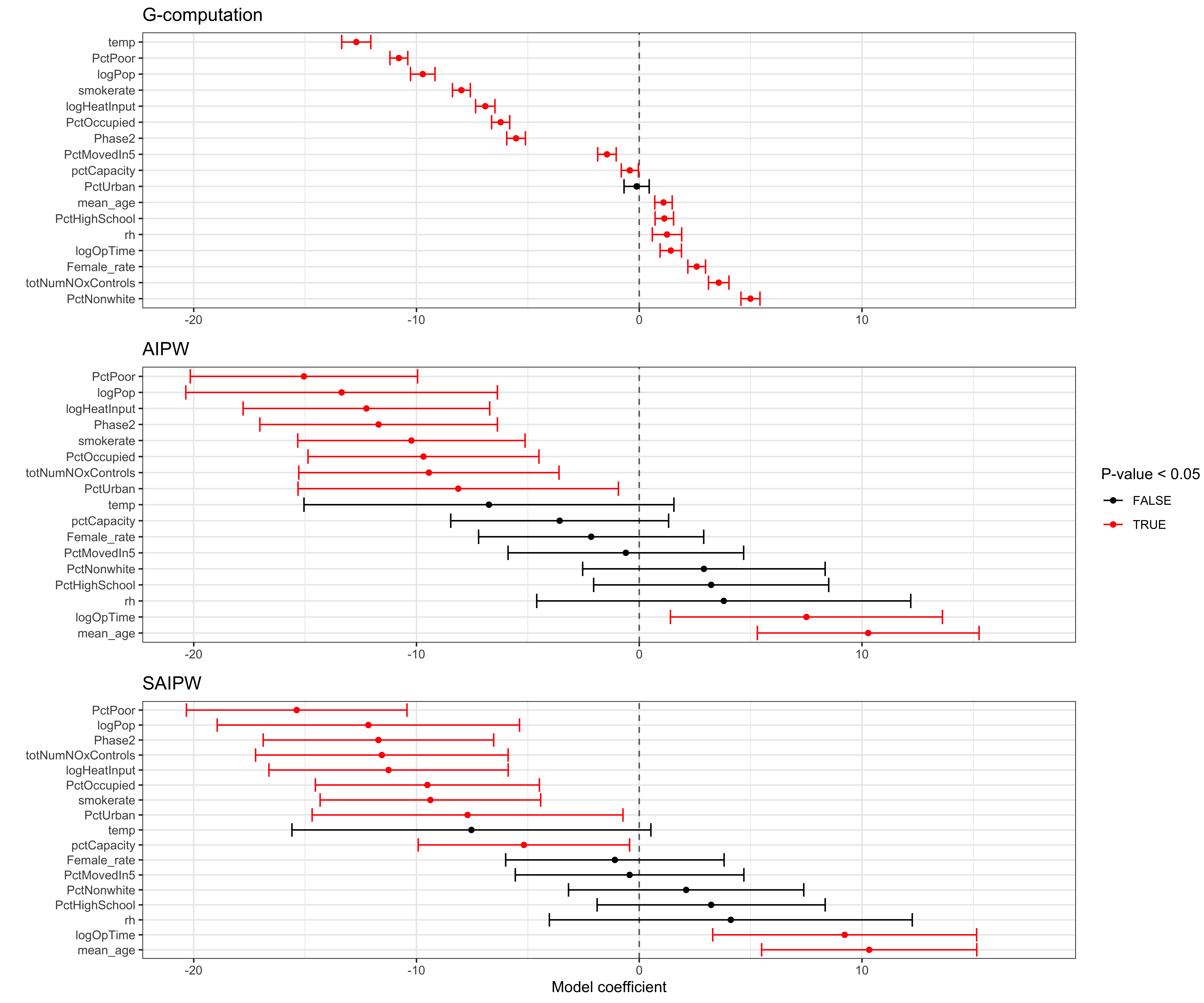}
    \caption{Heterogeneity of direct effects, holding upwind treatment at G=0, with the top and bottom 5\% of outcomes trimmed.}
    \label{fig:de_g0_05_trim}
\end{figure}

\begin{figure}[ht!]
    \centering
    \includegraphics[width=\linewidth]{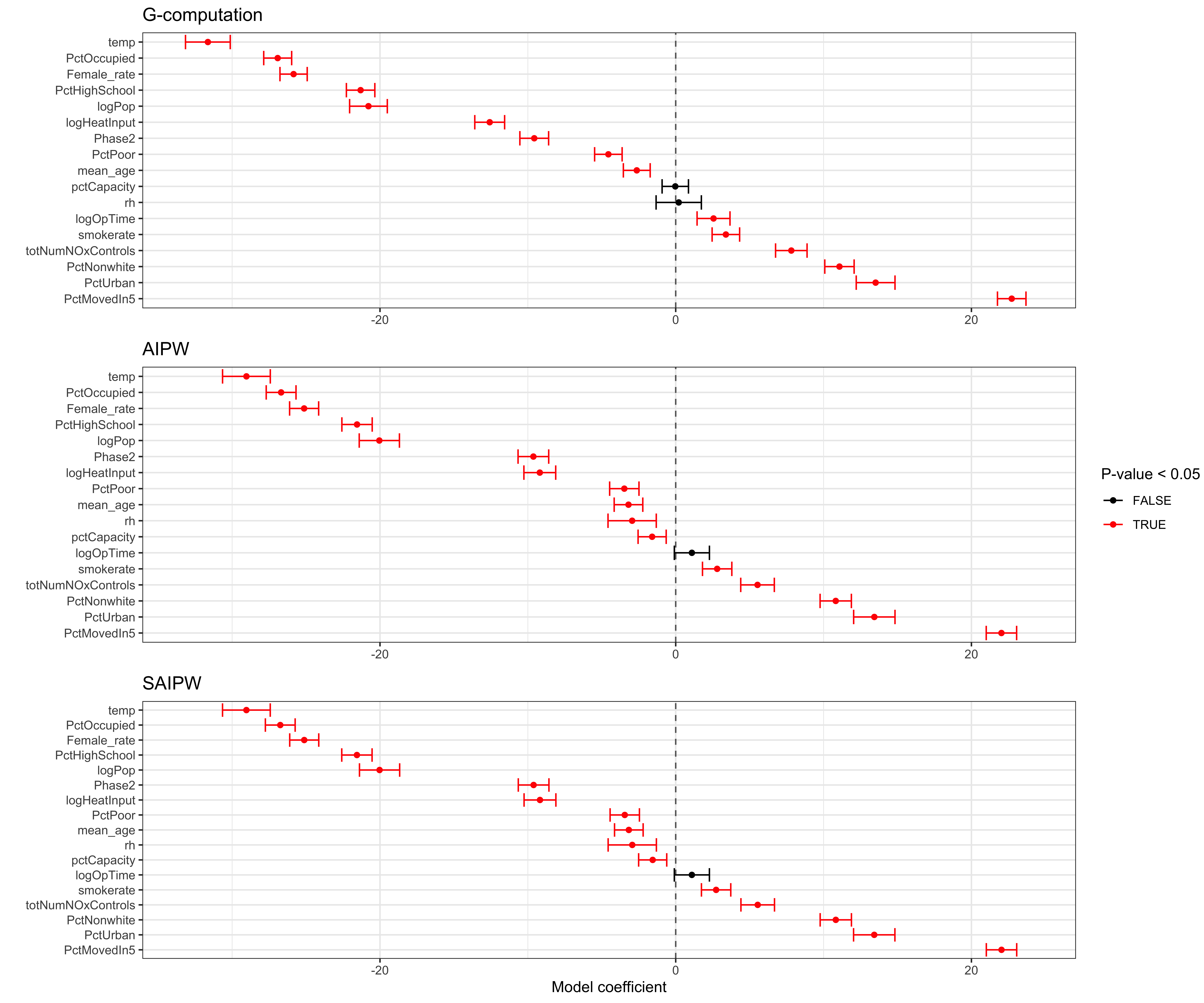}
    \caption{Heterogeneity of direct effects, holding upwind treatment at G=1.}
    \label{fig:de_g1_notrim}
\end{figure}

\begin{figure}[ht!]
    \centering
    \includegraphics[width=\linewidth]{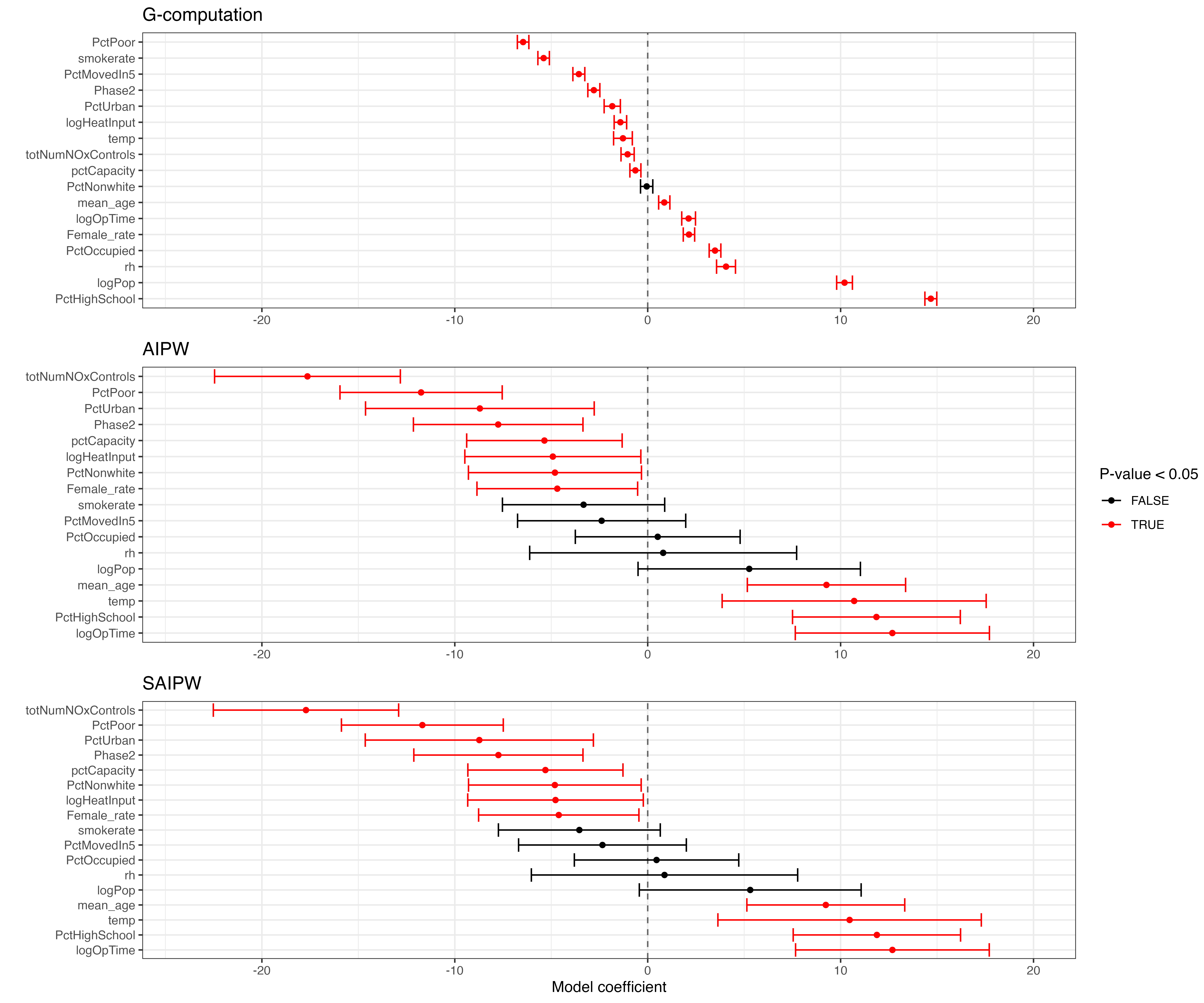}
    \caption{Heterogeneity of spillover effects, holding key-associated treatment at Z=1.}
    \label{fig:se_z0_notrim}
\end{figure}

\begin{figure}[ht!]
    \centering
    \includegraphics[width=\linewidth]{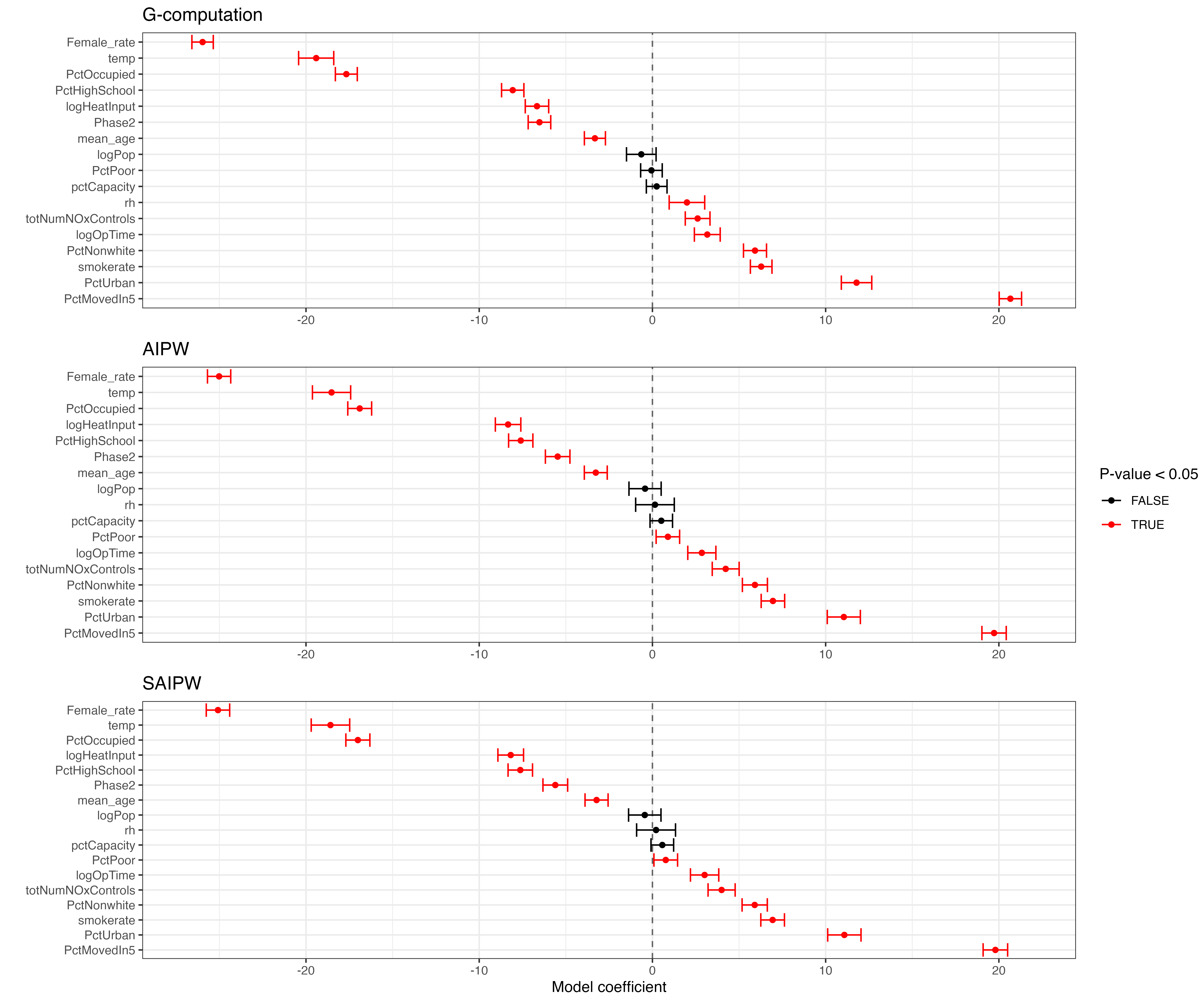}
    \caption{Heterogeneity of spillover effects, holding key-associated treatment at Z=1.}
    \label{fig:se_z1_notrim}
\end{figure}

%TC:endignore

% Sketch I:

% We are interested in 
% $$ E[\hat{\mu}^{AIPW}(A)] = E[AY/\hat{\phi}] + E[(1-A/\hat{\phi})\hat{\mu}] = I + II $$ 
% $$ I= E[\frac{1}{\hat{\phi}}E[AY \mid X]] =E[\frac{1}{\hat{\phi}}E[Y \mid X]E[A \mid X]] \mbox{ by }A \perp Y \mid X $$
% $$ II = E[(1-A/\hat{\phi})\hat{\mu}]=E[\hat{\mu}]-E[A/\hat{\phi} \hat{\mu}]=E[\hat{\mu}]-E[\frac{\hat{\mu} }{\hat{\phi}}E[A \mid X]]$$

% If $\hat{\phi}$ is correctly specified, then asymptotically, $$ I+II = E[Y]+E[\hat{\mu}]-E[\hat{\mu}]=E[Y] $$

% If $\hat{\mu}$ is correctly specified, then asymptotically, $$ I+II = E[\frac{1}{\hat{\phi}}E[Y \mid X]E[A \mid X]] +E[Y] - E[\frac{1}{\hat{\phi}} E[A Y \mid X]] =E[Y]$$
% %\addtolength{\textheight}{.5in}%

\end{document}